\documentclass[aps,prx,twocolumn,notitlepage,showpacs,amsmath,amstex,amssymb,citeautoscript,longbibliography,nofootinbib]{revtex4-1}
\pdfoutput=1
\usepackage{natbib}
\usepackage[english]{babel}
\usepackage{letltxmacro}
\usepackage{latexsym}
\usepackage{soul}
\usepackage{floatrow}
\LetLtxMacro{\ORIGselectlanguage}{\selectlanguage}
\makeatletter
\DeclareRobustCommand{\selectlanguage}[1]{%
  \@ifundefined{alias@\string#1}
    {\ORIGselectlanguage{#1}}
    {\begingroup\edef\x{\endgroup
       \noexpand\ORIGselectlanguage{\@nameuse{alias@#1}}}\x}%
}
\newcommand{\definelanguagealias}[2]{%
  \@namedef{alias@#1}{#2}%
}
\makeatother
\definelanguagealias{en}{english}
\definelanguagealias{English}{english}
\usepackage{graphicx}
\usepackage{amsmath}
\usepackage{amsfonts}
\usepackage{amssymb}
\usepackage{bm}
\usepackage{color}
\usepackage{soul}
\usepackage{wasysym}
\usepackage{dsfont}
\usepackage{bbold}
\usepackage{physics}
\usepackage{hyperref}
\hypersetup{
    bookmarks=false,         
    unicode=false,          
    pdftoolbar=false,        
    pdfmenubar=true,        
    pdffitwindow=false,     
    pdfstartview={FitH},    
    pdftitle={Particle density mobility edge},    
    pdfauthor={P. Brighi,  D.A. Abanin, M. Serbyn},     
    pdfsubject={},   
    pdfcreator={},   
    pdfproducer={}, 
    pdfkeywords={many-body localization} {mobility edge} {disordered systems}, 
    pdfnewwindow=true,      
    colorlinks=true,       
    linkcolor=black,          
    citecolor=blue,        
    filecolor=magenta,      
    urlcolor=blue           
}
\begin{document}
\title{Stability of mobility edges in disordered interacting systems}
\author{Pietro Brighi$^1$}
\author{Dmitry A. Abanin$^2$}
\author{Maksym Serbyn$^1$}
\affiliation{$^1$IST Austria, Am Campus 1, 3400 Klosterneuburg, Austria}
\affiliation{$^2$Department of Theoretical Physics, University of Geneva, 24 quai Ernest-Ansermet, 1211 Geneva, Switzerland}
\date{\today}
\begin{abstract}
Many-body localization provides a mechanism to avoid thermalization in isolated interacting quantum systems. The breakdown of thermalization may be complete, when all eigenstates in the many-body spectrum become localized, or partial, when the so-called many-body mobility edge separates localized and delocalized parts of the spectrum. Previously, De Roeck {\it et al.}~\href{https://dx.doi.org/10.1103/PhysRevB.93.014203}{[Phys. Rev.~{B} 93, 014203 (2016)]} suggested a possible instability of the many-body mobility edge in energy density. The local ergodic regions --- so called ``bubbles'' --- resonantly spread throughout the system, leading to delocalization. In order to study such instability mechanism, in this work we design a model featuring many-body mobility edge in \emph{particle density}:  the states at small particle density are localized, while increasing the density of particles leads to delocalization.  Using numerical simulations with matrix product states we demonstrate the stability of MBL with respect to small bubbles in large dilute systems for experimentally relevant timescales.  In addition, we demonstrate that processes where the bubble spreads are favored over processes that lead to resonant tunneling, suggesting a possible mechanism behind the observed stability of many-body mobility edge. We conclude by proposing experiments to probe particle density mobility edge in Bose-Hubbard model. 
\end{abstract}

\maketitle
\emph{Introduction.---}Many-body localization (MBL) provides a mechanism to avoid thermalization in isolated quantum interacting systems~\cite{Anderson80,Basko06}. Despite intensive theoretical~\cite{Abanin19,Huse-rev} and experimental~\cite{Bloch15,Bloch16,Bloch16-2,Google2019,Bloch2019,Kohlert19,Guo2019} studies, only fully-MBL phase in one spatial dimension is relatively well understood. The fate of MBL in higher dimensions~\cite{Roeck17,Wahl19,Dragos19,Eisert2019,Mirlin2020} and the possibility of the coexistence of localized and delocalized eigenstates in the same many-body spectrum~\cite{DeRoeck2016} remain debated.

Similarly to the case of Anderson localization~\cite{Mott67}, the MBL and delocalized eigenstates cannot coexist at the same energy suggesting the existence of \emph{many-body mobility edge (MBME)} --- a certain energy in the spectrum separating localized and delocalized eigenstates~\cite{Basko06}. In contrast to the non-interacting case, the energy of MBME scales \emph{extensively} with system size. In the absence of a coupling to a bath, this leads to an exactly vanishing conductivity (in contrast to an exponentially small but finite value in Anderson insulator) until a certain critical temperature~\cite{Basko06}.

Recently \textcite{DeRoeck2016} suggested a possible mechanism that may destroy MBME in large systems: a finite region with local energy density above the mobility edge --- a ``bubble'' --- may resonantly spread throughout the system thereby destroying localization everywhere. However, recent experiments~\cite{Guo2019} and MPS simulations~\cite{Chanda2020} gave evidence of MBME, at least on intermediate timescales. In addition, a number of numerical studies observed a mobility edge~\cite{Serbyn15,Alet14,Geraedts17,Goihl19,Yao} using exact diagonalization (ED). Unfortunately, the ED is limited to relatively small system sizes; experiments with MBME in energy density are also challenging since they require energy resolution.

In order to overcome the above challenges, we propose to study MBME \emph{in particle density}.  This allows us to directly probe the mechanism of instability suggested in Ref.~\cite{DeRoeck2016}, which equally applies to MBME in any extensive conserved quantity. First, using numerical simulation with matrix product states~(MPS), we demonstrate that uniform dilute states remain localized even at system sizes of $L=40$ sites up to 250 tunneling times (i.e.\ more than two orders of magnitude larger than the inverse local hopping). Next, we use a region with large particle density to reproduce the bubble described in~\cite{DeRoeck2016} and track its influence on the dilute remainder of the system in a quantum quench.  We do not find any evidence of resonant tunneling of the bubble, at least on experimentally relevant timescales.

In summary, the study of the particle density MBME facilitates the state preparation and analysis and allows us to access the dynamics of much larger systems using time evolution with MPS. We report the stability of the particle density mobility edge on long timescales and suggest that similar physics may be experimentally probed using Bose-Hubbard model.

 \emph{Correlated hopping model.---}We consider hard-core bosons on an open chain of size $L$, with the following Hamiltonian,
\begin{multline}
\label{Eq:H}
\hat{H} = t_1\sum_{i=1}^{L-1} (c^\dagger_{i+1}c_i + {\rm h.c.}) + \sum_{i=1}^L \epsilon_i \hat{n}_i \\
 + t_2\sum_{i=2}^{L-1}(c^\dagger_{i-1}\hat{n}_i c_{i+1} + {\rm h.c.}).
\end{multline} 
The first line corresponds to the non-interacting Anderson's model \cite{Anderson58}, where random on-site potential has a uniform distribution, $\epsilon_i\in[-W,W]$. The facilitated hopping in the second line enables motion of a \emph{pair of particles} with amplitude $t_2$, $ {\bullet}{\bullet}{\circ}{\leftrightarrow}{\circ}{\bullet}\bullet$, making the model interacting. The Hamiltonian~(\ref{Eq:H}) has two channels for dynamics: the single particle hopping prevails in dilute states, while the pair hopping is dominant at larger densities. 

We note that a similar model was discussed in Ref.~\cite{DeRoeck2016} in two dimensions, although only {with} two particles. The enhancement of localization length in the case of two interacting particles also received significant attention~\cite{Shep1994,Imry1995}. In a different direction, the fate of the single particle mobility edge in the presence of interactions was studied~\cite{li2015,Kohlert19}. In contrast, we study model~(\ref{Eq:H}) that does not have a single particle mobility edge and consider the finite particle density regime.

We fix the value of the hopping parameters $t_1 = 0.5$ and $t_2=2$ so that  the localization length of a single particle $\xi_\text{SP} \lesssim 1$ and at the same time a single pair has a localization length $\xi_\text{P} \gtrsim  2.5$ for $2.5\lesssim W\lesssim 6$~\cite{SOM}. For such a choice, our model does not suffer from finite size effects~\cite{Papic15} and we establish MBME using eigenstates probes.

\emph{Eigenstate probes of localization.---}We use exact diagonalization and shift-invert (SI) numerical techniques to provide evidence for MBME in Hamiltonian~(\ref{Eq:H}). We analyze the average ratio of level spacings, $\delta_i = E_{i+1}-E_i$, in the middle of the spectrum, $r_\text{av}  = \langle\min (\delta_i,\delta_{i+1})/\max (\delta_i,\delta_{i+1})\rangle$. This is a commonly used probe of the MBL transition~\cite{PalHuse,Alet14} that attains the value $r_\text{P} \simeq0.39$ for the Poisson level statistics, characteristic of the MBL phase, and $r_\text{GOE}\simeq 0.53$ for the case of random Gaussian orthogonal ensemble (GOE), typical for chaotic Hamiltonians with time-reversal symmetry.  

\begin{figure}[t]
\begin{center}
\includegraphics[width=0.95\columnwidth]{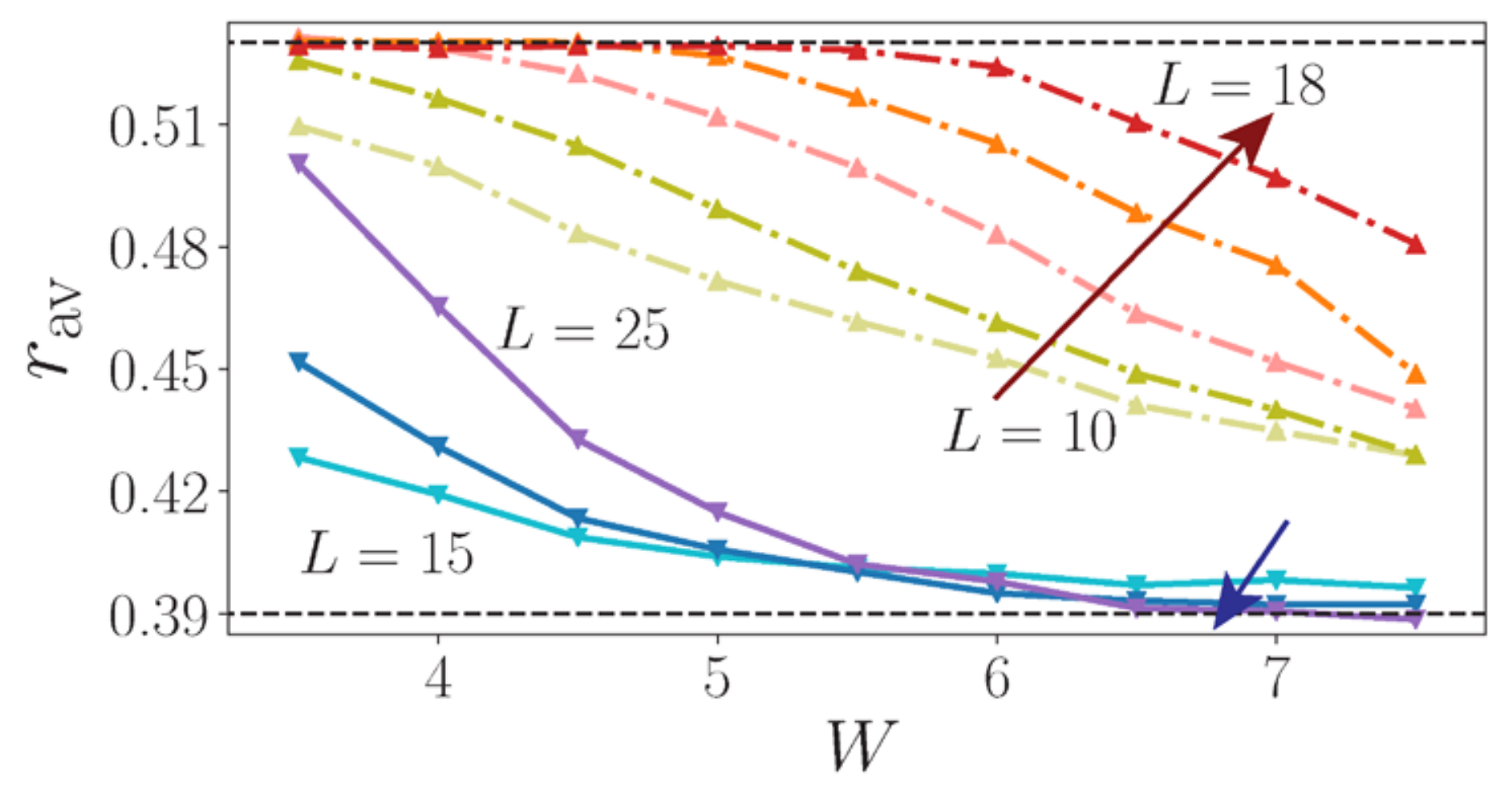}\\
\caption{Scaling of level spacing ratio demonstrates that at density $\nu=1/5$ (solid lines, $L=15,20,25$ with $3,4,5$ particles) the system enters MBL phase for $W\geq 6.3$. In contrast, at half-filling $\nu=1/2$ (dashed curves, $L=10,\ldots 18$) the critical disorder strengths is much larger and in the entire range of disorder $r_\text{av}$ approaches thermal value with increasing system size. Data is generated from ED/SI simulations with at least $10^3$ disorder realizations using approximately $2\%$ of eigenstates in the center of the spectrum.\label{Fig:r-ratio}}
\end{center}
\end{figure}

Figure~\ref{Fig:r-ratio} displays that at half-filling, $\nu = N/L=1/2$, where $N$ is the total number of particles and $L$ is the chain length, the level statistics approaches GOE with increasing system size, which is consistent with the delocalized phase. In contrast, at $\nu=1/5$ filling $r_\text{av}$ flows towards $r_\text{P}$ at strong disorder. In what follows we fix the disorder strength to be $W=6.5$, since at this value the dilute limit is localized while the dense limit clearly flows towards delocalization. The scaling of entanglement entropy also confirms the coexistence of localized and delocalized phases for disorder $W=6.5$~\cite{SOM}. 

\begin{figure*}
\includegraphics[width=0.95\columnwidth]{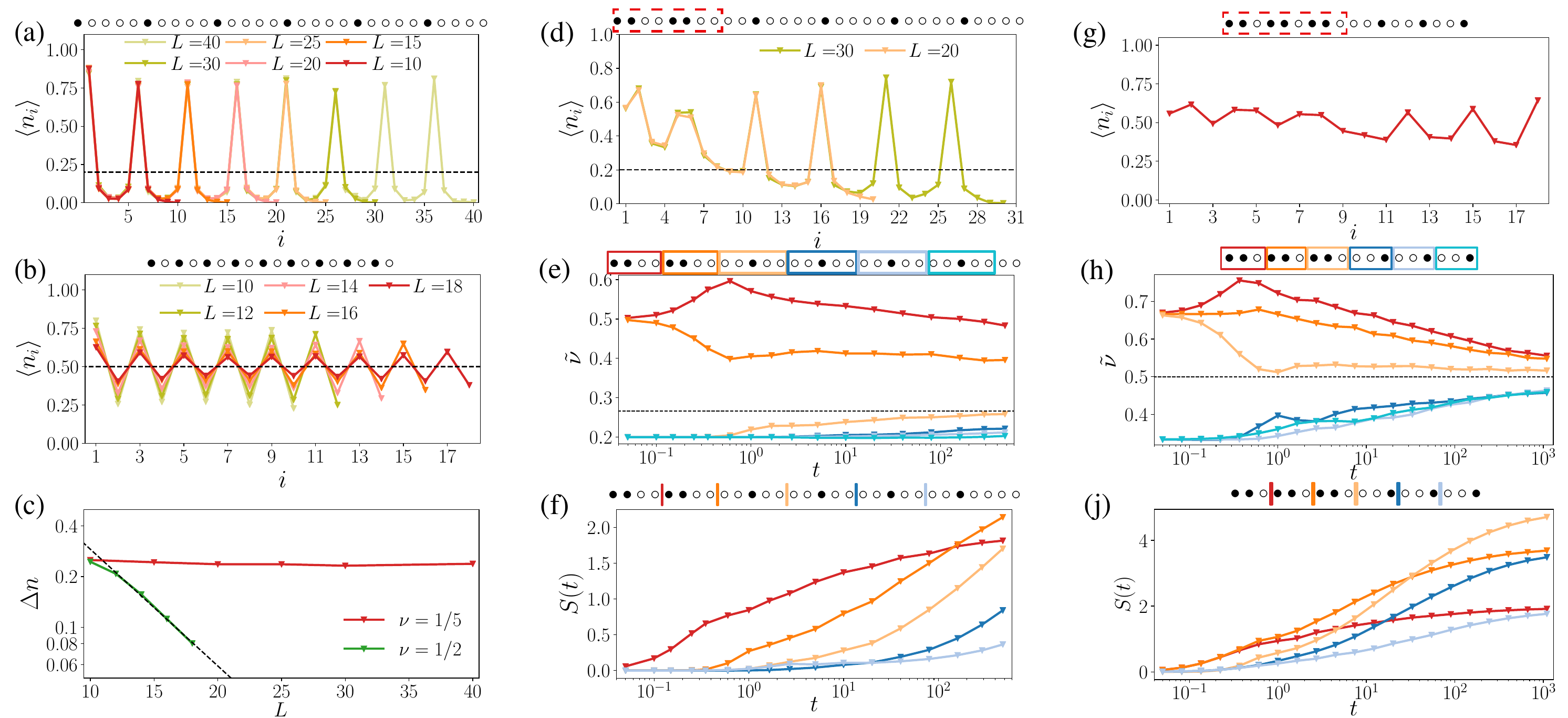}
\caption{\label{Fig:quench} 
\textbf{(a)-(c)} The quantum quench from the uniform density wave with period $1/\nu$ reveals  memory of the initial state at $\nu=1/5$ in (a), whereas in the dense case $\nu=1/2$ (b) the charge pattern relaxes to zero exponentially in the system size as is shown in (c). \textbf{(d)-(f)} Stability of the dilute system against the bubble consisting of a half-filled region with 4 particles is illustrated in panel (d) by the density profile at $T_\text{max}$. (e) The time dynamics of density in the coarse grained regions (see the legend at the top) shows the absence of significant relaxation in regions away from the bubble. The density in the region at the boundary with the bubble increases logarithmically in time. (f) The onset of logarithmic entanglement dynamics after a transient is visible for all cuts (see the legend at the top) away from the bubble.
 \textbf{(g)-(j)} In contrast, the bubble delocalizes the system when the overall density $\nu=1/2$. The residual density profile at $T_\text{max}=1000$ in panel (g) has only weak memory of the initial state. In addition, densities coarse grained over 3-site regions in (f) all tend to the equilibrium value of 1/2 and entanglement entropy in (j) displays faster than logarithmic growth for all cuts. The data are generated using TEBD and Krylov (ED) dynamics using between $5\times 10^4$--$100$ (dilute) and $3\times 10^4$--$10^3$ (dense) disorder realizations.
}
\end{figure*}

\emph{Quench dynamics.---}Having provided numerical evidence for the coexistence of localized and delocalized phases in small systems, we turn to quantum quench dynamics that distinguishes MBL from ergodic phase~\cite{Serbyn14,Bloch15}. We consider quenches where the system is initially prepared in a product state and then evolved with the Hamiltonian~(\ref{Eq:H}). Starting with a density wave of period $1/\nu$, a configuration that contains no pairs, we calculate the density profile at late times. For the dilute case, $\nu=1/5$, we use the time-evolved block decimation (TEBD) with MPS~\cite{tebd,ITensor} (see~\cite{SOM} for additional details and benchmarks). This allows to monitor dynamics of systems as large as $L=40$ sites up to times $T_\text{max}=500$.  In the dense case ($\nu=1/2$) we use ED and Krylov subspace time evolution method. While ED allows to access the infinite-time density profile, with the Krylov method, we simulate quantum dynamics up to $T_\text{max}=1000$. 

The density profiles at late times look very different in the dense and dilute cases. While in the dilute case the system retains memory of the initial state, see Fig.~\ref{Fig:quench}(a), at $\nu=1/2$ quantum dynamics leads to a progressively more uniform density profile with increasing system size, Fig.~\ref{Fig:quench}(b). In order to quantify the difference in the form of the density profile at late times, in Fig.~\ref{Fig:quench}(c) we plot the average deviation of the density from the equilibrium thermal value $\nu$, $\Delta n=({1}/{L})\sum_{i=1}^L|\langle \hat{n}_i(T_\text{max}) \rangle -\nu|$. The deviation of late-time density from the thermal value, $\Delta n$, in the dense regime decays exponentially with the system size as $\Delta n\sim e^{-L/\xi_\text{T}}$, where $\xi_\text{T} \simeq 6.27$. In contrast, for the dilute case $\Delta n$ shows no dependence on the system size, as is apparent in the density profiles. The characteristic length $\xi_\text{T}$ extracted in the dense case gives the minimum size for genuine egodic bubbles that can destroy the MBME according to Ref.~\cite{DeRoeck2016}. 

Having confirmed the coexistence of localized and delocalized states at different values of particle density $\nu$ for the same disorder strength, we proceed with a more detailed study of the effect of a bubble, whose behavior is central to the mechanism proposed in~\cite{DeRoeck2016}. Figure~\ref{Fig:quench}(d) illustrates the evolution of a non-uniform initial state, where a dense region represents the bubble. The bubble region consists of $8$ sites with two pairs of particles and has a local density of $\nu=1/2$. The bubble is followed by a period-5 density wave that occupies $L-10$ sites and two additional empty sites at the end of the chain. Although having $\nu=8/30>1/5$, this state is still in a localized sector, as shown in~\cite{SOM}. The bubble leaks only weakly into the dilute region even at late times, see Fig.~\ref{Fig:quench}(d), with particles far away from the bubble not being affected. In contrast, in the dense case, Fig.~\ref{Fig:quench}(g), the bubble with average density of $\nu=2/3$ successfully melts the period-3 density wave state throughout the system. 

Next, in panels Fig.~\ref{Fig:quench}(e) and (h) we further illustrate the differences between the density dynamics in the dense and dilute cases in presence of a bubble. In both cases we plot the density of particles within subregions of small size $k$, $\tilde \nu_i = (1/k)\sum_{j=i}^{i+k-1} \langle n_j\rangle$, that are shown at the top of the plot. In the dilute case, Fig.~\ref{Fig:quench}(e), we observe that $\tilde \nu$ remains far from its thermal value even at late times, in contrast with~\cite{DeRoeck2016}, where an ergodic region larger than $\xi_T$ is expected to delocalize the system. The densities of regions in the bubble and adjacent to the bubble seem to saturate, while the regions far away from the bubble show very slow dynamics. In contrast, the dense case, Fig.~\ref{Fig:quench}(h), shows that all expectation values evolve towards equilibrium, although the regions far away from the center of the chain display slow, logarithmic in time,  growth of density. 

Finally, we study the dynamics of the bipartite entanglement entropy, $S_\text{vN}$, see Fig.~\ref{Fig:quench}(f) and (j).  The entanglement is defined as $S_\text{vN}= -\tr \rho \ln \rho$, where $\rho$ is the density matrix of the left subregion calculated from $\ket{\psi(t)} = e^{-i\hat Ht}\ket{\psi_0}$. Different entanglement cuts shown at the top of Fig.~\ref{Fig:quench}(f) and (j) are encoded by their color. Consistent with MBL, the increase of entanglement in the region close to the bubble is logarithmic in time in Fig.~\ref{Fig:quench}(f)~\cite{Znidaric08,Moore12,Serbyn13-2,lukin2018}.  The entanglement across the cuts further away from the bubble begins to grow at significantly later times. For these more distant cuts, the initial uprise in entanglement corresponds to a slow logarithmic change of density~[see Fig.~\ref{Fig:quench}(e)], and after saturation of density dynamics, we expect an onset of the logarithmic growth of entanglement. In contrast, the entanglement dynamics in Fig.~\ref{Fig:quench}(j) is always faster than logarithmic. In \cite{SOM} we provide more details on the contribution of particle transport to entanglement~\cite{islam2015,lukin2018}, demonstrating that it is responsible for logarithmic entanglement increase, in agreement with~\cite{Kiefer-Emmanoulidis2020}, whereas the configurational entanglement grows faster than logarithmic, and total entropy shows power-law increase.

\noindent\emph{Bubble tunneling vs.~decay processes.---}The quench dynamics discussed above suggests that a bubble is not able to spread through the entire localized chain and remains in the vicinity of its initial position. At the same time, most of our quench simulations are restricted to finite, albeit long, times. In order to give a complementary evidence for the bubble localization, we return to eigenstate properties that effectively probe the infinite time limit.  

\begin{figure}[b]
\begin{center}
\includegraphics[width=0.95\columnwidth]{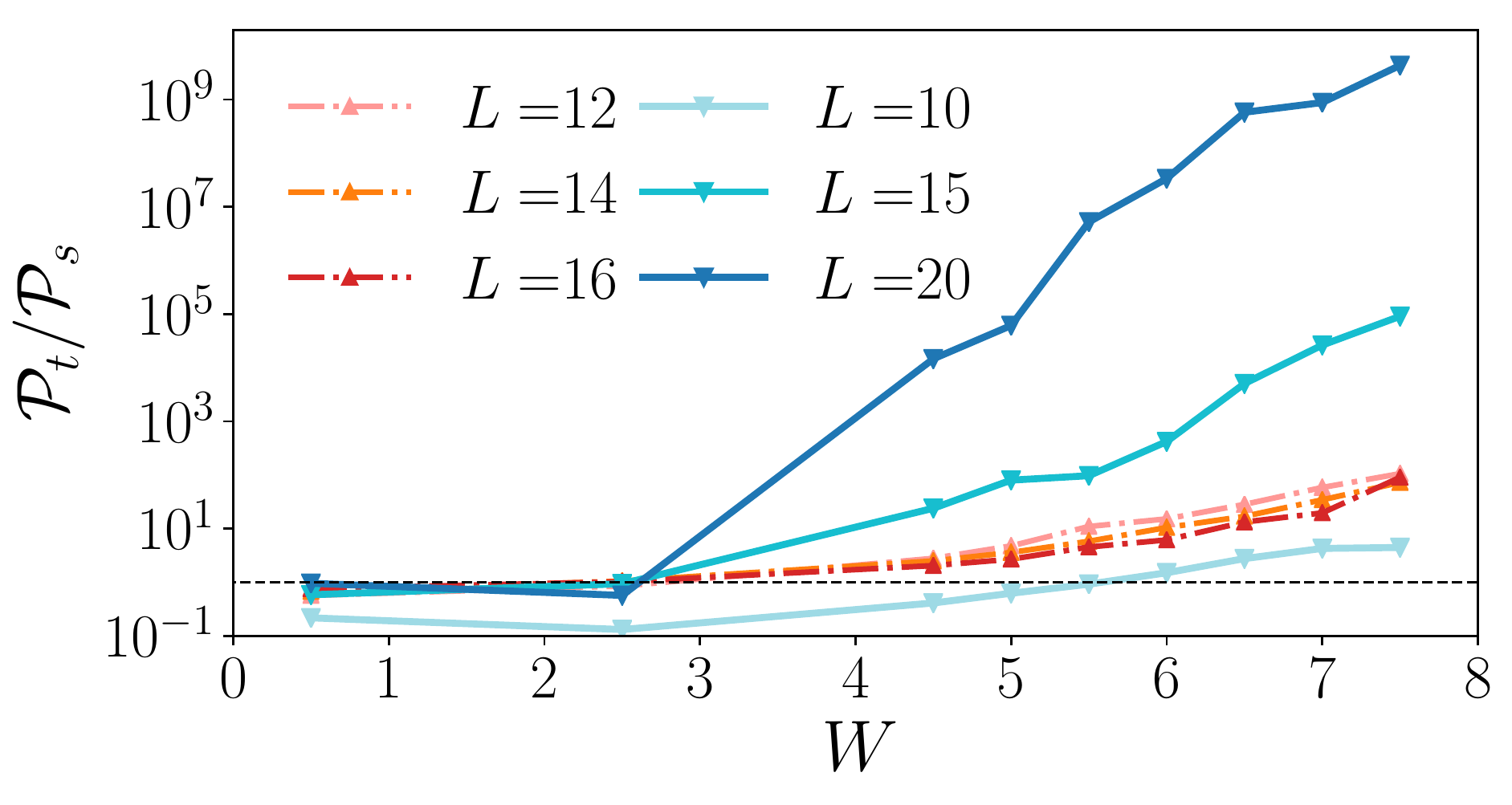}\\
\caption{ The rapid increase of the ratio of ${\cal P}_\text{s}/{\cal P}_\text{t}$ with system size and disorder strength reveals that in the dilute case, $\nu=1/5$ the probability for bubble spreading is strongly enhanced compared to the probability of bubble tunneling to the opposite end of the system. For the dense case these two probabilities are of the same order and approach each other with increasing system size in a broad range of disorders. Averaging is done over at least $2.5\times 10^3$ disorder realizations.
\label{Fig:mutual-iprs}}
\end{center}
\end{figure} 

We start with an initial product state in the half-filled case illustrated for $L=12$,
\begin{equation}\label{Eq:psi-bubble}
\ket{\psi_1}=\boxed{{\bullet}\bullet\circ\bullet{\bullet}\,{\circ}}\circ\circ\bullet\circ\circ\bullet,
\end{equation}
that contains a bubble of $k=L/2$ sites with $\nu=2/3$ filling (boxed region), followed by a sparser region with the same number of sites and density $\nu  = 1/3$. To quantify the relation between  the probability of the bubble tunneling to the opposite end of the system and the probability of the bubble spreading throughout the system, we use a spatial reflection of $\ket{\psi_1}$ and uniform density wave as a representative of the state with bubble tunneling and spreading, respectively: 
\begin{eqnarray}\label{Eq:psi-inverted}
\ket{\psi^\text{t}_2}&=&\bullet\circ\circ\bullet\circ\circ\boxed{\circ\bullet\bullet\circ\bullet\,\bullet},\\
\label{Eq:psi-spread}
\ket{\psi^\text{s}_2}&=&\bullet\circ\bullet\circ \bullet\circ\bullet\circ\bullet\circ\bullet\circ\bullet,
\end{eqnarray} 
illustrated for $L=12$ and $\nu=1/2$ filling.  For dilute configurations at $\nu=1/5$ we define the bubble as a region of size $2(\nu L-1)$ with density $\nu=1/2$, joined with a dilute remainder. For $L=20$ such a  state is:\\ $\ket{\psi_1}=\boxed{{\bullet}\bullet\circ\bullet{\circ}\,{\circ}}\circ\circ\circ\circ\circ\circ\circ\bullet\circ\circ\circ\circ\circ\circ\circ\circ$.

It is straightforward to show that the infinite time average probability of finding the system with the wave function $e^{-i\hat H t}\ket{\psi_1}$ in the product state $\ket{\psi_2}$ is given by 
\begin{equation}
    \label{Eq:MutIPR}
    \mathcal{P}(\ket{\psi_1},\ket{\psi_2})=\sum_{\alpha=1}^\mathcal{N}|\bra{E_\alpha}\ket{\psi_1}\bra{E_\alpha}\ket{\psi_2}|^2,
\end{equation}
where $\ket{E_\alpha}$ are the complete set of eigenstates of $\hat H$. Eq.~(\ref{Eq:MutIPR}) quantifies the similarity in the expansion of two different states  $\ket{\psi_{1,2}}$ over the basis $\ket{E_\alpha}$ and reduces to the conventional participation ratio when $\ket{\psi_1}=\ket{\psi_2}$.

In order to reveal the relation between bubble decay and tunneling processes, we calculate the ratio of probabilities of bubble decaying, ${\cal P}_\text{s} = {\cal P}(\ket{\psi_1},\ket{\psi^\text{s}_2})$,  with $\ket{\psi^\text{s}_2}$ from Eq.~(\ref{Eq:psi-spread}),   to bubble tunneling, ${\cal P}_\text{t} = {\cal P}(\ket{\psi_1},\ket{\psi^\text{t}_2})$ with $\ket{\psi^\text{t}_2}$ from Eq.~(\ref{Eq:psi-inverted}). In the dense case, these two probabilities are of the same order and moreover tend to identity with increasing system size as expected in the delocalized phase, see Fig.~\ref{Fig:mutual-iprs}. In the dilute case, the ratio ${\cal P}_\text{s}/{\cal P}_\text{t}$ is rapidly increasing with both disorder, and system size. This  proves that the bubble tunneling processes are strongly suppressed compared to the processes where the bubble spreads throughout the system, calling into question the applicability of the resonance argument of~\cite{DeRoeck2016}.

\emph{Experimental realization.---} Finally, we discuss a possible way to observe the physics related to MBME in experiments with ultracold atoms. Within the disordered Aubry-Andr\'e bosonic Hamiltonian,
\begin{equation}\label{Eq:Ham2}
\hat H =  \sum_{i} \left[ t (a_{i}^\dagger  a_{i+1} + \text{h.c.})+ \epsilon_i  n_{i,\sigma} + U n_{i}  (n_{i}-1)\right],\end{equation}
 that is actively used to study MBL physics~\cite{lukin2018,Rispoli2019}, the bubbles can be represented by regions with $\langle a_i^\dagger a_i\rangle = \rho>1$ bosons per site. A particle within such region has a hopping matrix element enhanced by the Bose-factor of $\langle\rho\rangle$, thus playing the role of hopping $t_2$ in model~(\ref{Eq:H}). In the regime of densities and disorder strengths such that the enhanced hopping $\langle\rho\rangle t$ corresponds to localization lengths significantly larger than lattice spacing, $\xi_\text{dense} > a$, whereas a single boson localization length is $\xi\lesssim a$, this model will implement similar physics to our toy model. Note that at the same time it is important to keep interaction $U$ low enough,  $U\leq t$, to avoid the formation of minibands related to long-lived doublons. 

By initializing the system in a product state with a dense region of bosons in the center of the trap along with low density of bosons away from such a region, the dynamics under Hamiltonian~(\ref{Eq:Ham2}) will probe the ability of the bubble to melt the imbalance~\cite{Bloch15} away from its original position. From our simulations we expect the absence of imbalance relaxation far away from the bubble. In a different direction, doublons~\cite{Krause19,Iadecola18} or second species of particles not subject to disorder~\cite{Bloch2019} are also promising candidates that can play a role of the bubble. 

\emph{Discussion.---}We presented a model with MBME in particle density and investigated its properties numerically using ED and time evolution with MPS. We find strong evidence of the persistence of localization at infinite times for small systems and also observe memory of initial configuration until times of $T_\text{max}=500$ for systems with up to $L=40$ sites. These times are at least two orders of magnitude larger compared to the inverse local hopping, $\hbar/t_1$, and are achievable with cold atoms experiments. While we cannot rule out a residual very slow delocalization at much later times, the constructed model allows us to bound the timescale up to which the localization remains stable in very large systems that are beyond the reach for ED. 

The model with MBME in particle density presented in this work allows for direct tests of the arguments about the instability of MBME~\cite{DeRoeck2016}. In order for bubble to move throughout the system it is important that the bubble does not disappear by spreading and that configurations with bubbles situated at different locations are effectively coupled to each other. Our simulations reveal that dilute systems have no trace of  bubble reemerging at a different location within the system. Moreover, even the expectation value of the pair density $\langle n_in_{i+1}\rangle$~(pairs are building blocks of the bubble) is exponentially suppressed away from the original location of the dense bubble~\cite{SOM}. In an alternative approach, we directly test the probability of the bubble to emerge at the opposite end of the system at infinite time and find it to be strongly suppressed. 

To conclude, we expect that the proposed model will enable further investigations of particle density MBME. Studies of the structure of matrix elements, extension of the theory of LIOMs~\cite{Serbyn13-1,Huse13} to systems with MBME in particle density~\cite{Geraedts17}, and studies of the effect of a small bath on a localized system~\cite{nandkishore2015many,Luitz17,hyatt2017,Crowley19} using our model represent promising avenues for future work.

\emph{Acknowledgments.---}We acknowledge useful discussions with W.~De Roeck and A.~Michailidis. P.B.~was supported by the European Unions Horizon 2020 research and innovation programme under the Marie Sklodowska-Curie Grant Agreement No.~665385. D.A.~was supported by the Swiss National Science Foundation. M.S.~was supported by European Research Council (ERC) under the European Union's Horizon 2020 research and innovation programme (grant agreement No.~850899)  This work benefitted from visits to KITP, supported by the National Science Foundation under Grant No. NSF PHY-1748958 and from the program ``Thermalization, Many body localization and Hydrodynamics'' at International Centre for Theoretical Sciences (Code: ICTS/hydrodynamics2019/11).

\clearpage
\pagebreak
\onecolumngrid

\begin{center}
\textbf{\large Supplementary material for ``Particle density mobility edge'' }\\[5pt]
\begin{quote}
{\small 
In this supplementary material we present additional data and details of the methods used in the main text. First, we show data for additional probes of ETH breakdown such as entanglement entropy. After this we present benchmarks of our time-evolving block decimation simulation of dynamics. Finally, we explore the behavior of the mutual IPR defined in the main text.
}\\[20pt]
\end{quote}
\end{center}
\setcounter{equation}{0}
\setcounter{figure}{0}
\setcounter{table}{0}
\setcounter{page}{1}
\setcounter{section}{0}
\makeatletter
\renewcommand{\theequation}{S\arabic{equation}}
\renewcommand{\thefigure}{S\arabic{figure}}
\renewcommand{\thepage}{S\arabic{page}}

\twocolumngrid

\section{Localization length and parameter choice}\label{Sec:loc-length}
\begin{figure}[b]
\begin{center}
\includegraphics[width=0.95\columnwidth]{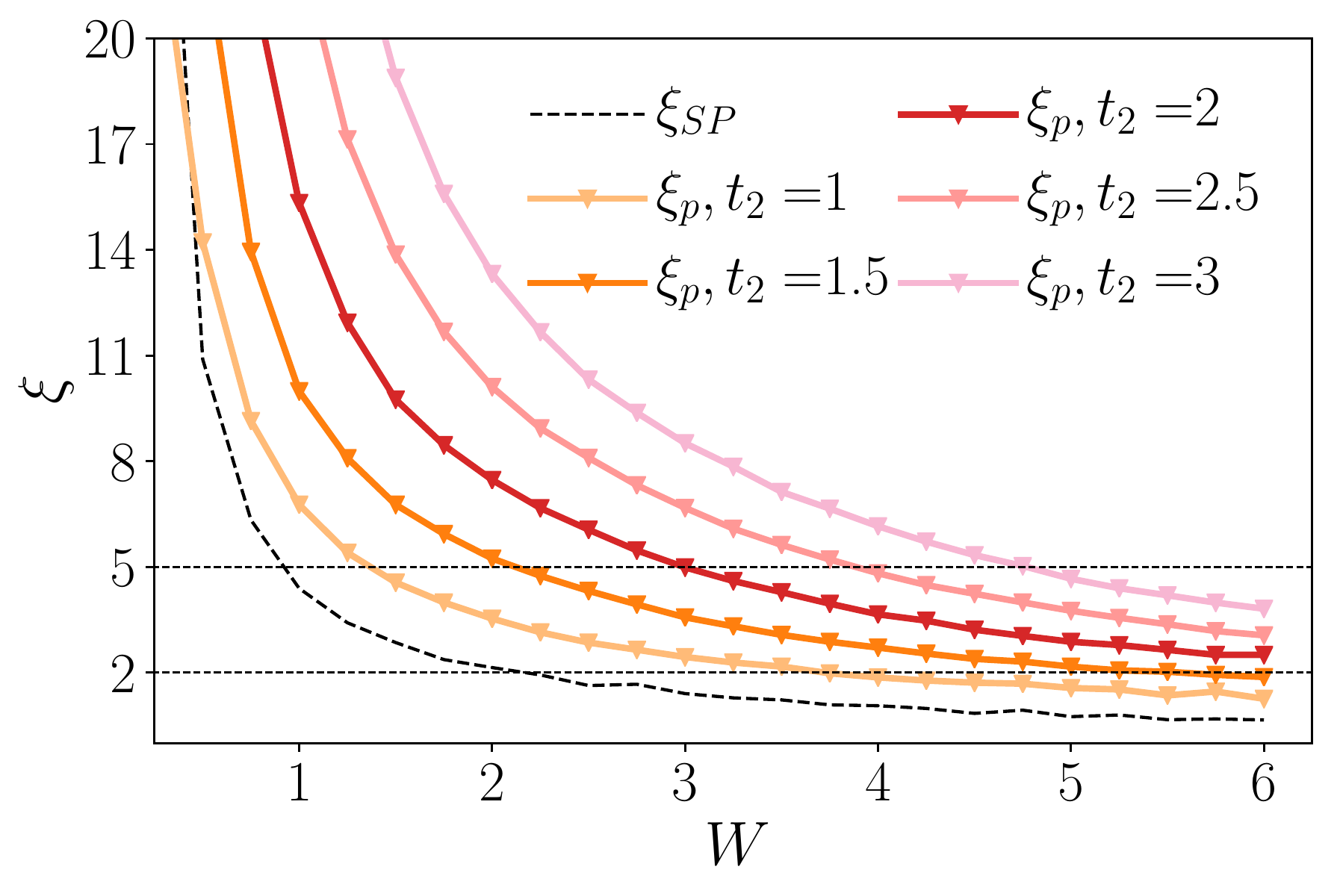}\\
\caption{The localization length decreases, as expected, with the disorder strength for all the values of $t_2$. For every constrained hopping amplitude $t_2$ it is possible to locate the region of disorder where we expect to see a MBME in particle density as the area among the two dashed lines. As the curve crosses the first dashed line, systems with typical particle spacing 5 will be localized. Nevertheless denser states will still be delocalized, having smaller distance among particles. Data were obtained on a lattice of length $L=50$ and averaged over $5000$ disorder realizations. \label{Fig:Loc-Length}}
\end{center}
\end{figure}

The dynamics generated by the constrained Hamiltonian, Eq.~(\ref{Eq:H}), strongly depends on the choice of the hopping parameters $t_{1,2}$. In order to choose the most suitable parameters for the study of MBME in particle density, we explore localization lengths for a single particle $\xi_\text{SP}$ and for one pair of particles $\xi_\text{P}$. These localization lengths are evaluated using ED. We calculate the infinite-time average of the occupation number at each site for an initial state where either a single particle or a single pair are initialized at the first site of the chain. We extract the localization lengths $\xi_\text{SP}$ ($\xi_\text{P}$) from an exponential fit of the density curve $\langle n_i \rangle$ . 

Resulting values of $\xi_\text{P,SP}$ for fixed $t_1=0.5$ and different disorder values and different values of hopping $t_2$ are shown in Fig.~\ref{Fig:Loc-Length}. The single particle hopping localization length (dashed line in Fig.~\ref{Fig:Loc-Length}) does not depend on $t_2$, and becomes smaller than one lattice spacing for $W\gtrsim 4$. The pair localization length is monotonously increasing with  $t_2$ at fixed value of disorder strength, $W$. Our aim is to have $\xi_\text{P}$ in the range between $2$ and $5$. In this regime, the half-filling case is expected to be delocalized, while at lower densities $\nu\sim 1/5$, when the typical distance between pairs is large, we expect MBL phase. This motivates the choice $t_2=2$, since at this value of $t_2$  $\xi_\text{P}(W)$ approaches $2$ at disorder strength around $W\sim 6$. We note that we avoided further increase of $t_2$ to keep the model away from the constrained limit: in the case when $t_2$ dominates over $t_1$, the model would approximately reduce to a kinematically constrained model that has many disconnected sectors in the Hilbert space. 

\begin{figure}[b]
\begin{center}
\includegraphics[width=0.95\columnwidth]{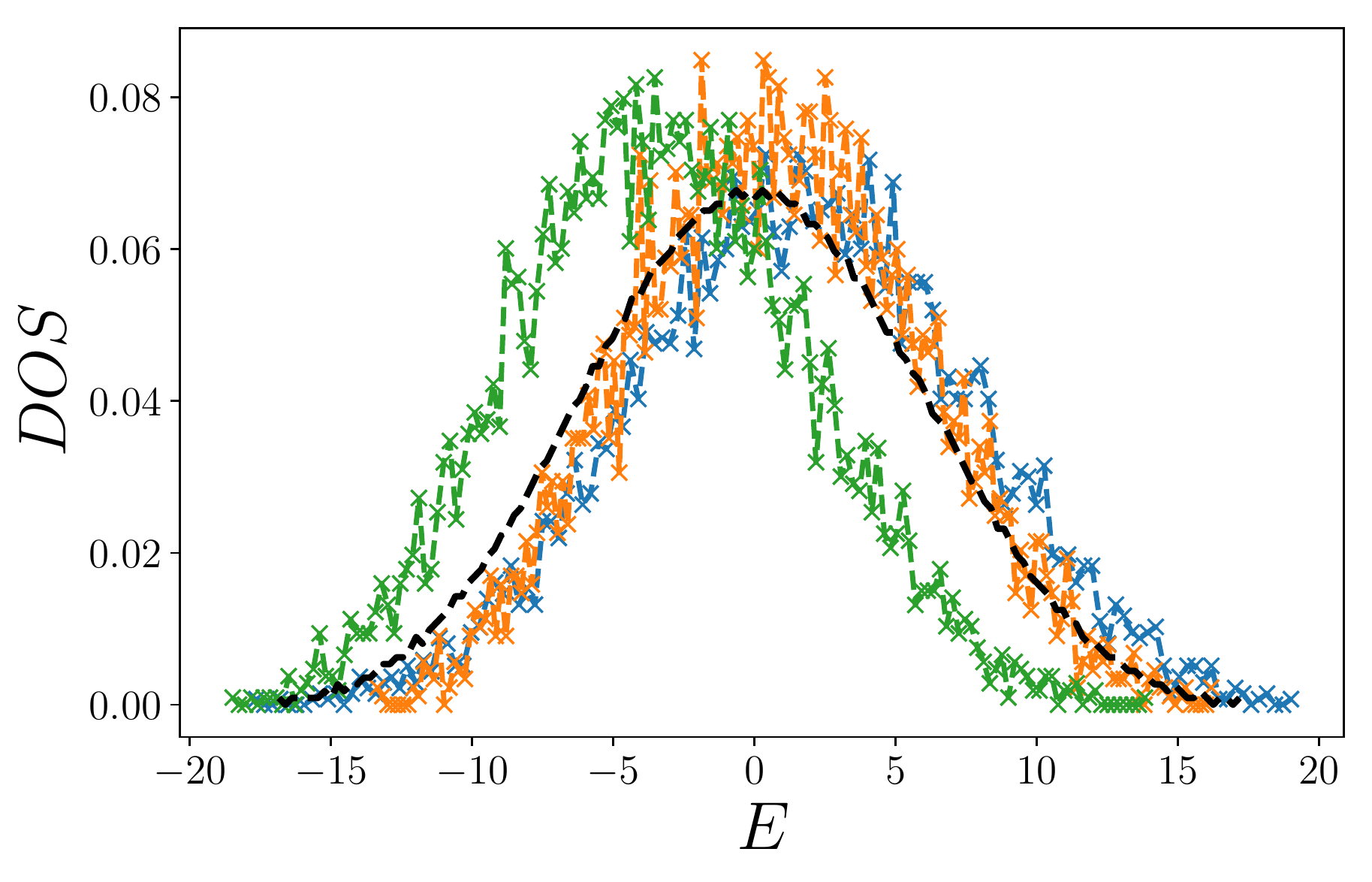}\\
\caption{The DOS from single disorder realizations show a relatively smooth behavior and a Gaussian shape, thus confirming the absence of strong finite size effects. DOS refers to a chain with $L=20$ and $\nu=1/4$. Disorder strength is $W=5.0$. Green, blue and orange curves correspond to different disorder realizations, while the black dashed line shows disorder-averaged DOS. \label{Fig:DOS}}
\end{center}
\end{figure}

In order to rule out the presence of strong finite size effects, we studied the density of state in individual disorder realizations. In the regime when $t_2\gg t_1$ the strong finite size effects would give rise to the presence of the mini-bands and the DOS would become non-monotonous with numerous peaks corresponding to mini-band structure~\cite{Papic15}. Figure~\ref{Fig:DOS} confirms that at our choice of parameters even individual disorder realizations have a relatively smooth density of states with Gaussian envelope, thus ruling out the presence of strong finite size effects. 

\section{ED probes of localization}\label{Sec:ED-probes}
\begin{figure}[tb]
\begin{center}
\includegraphics[width=.95\columnwidth]{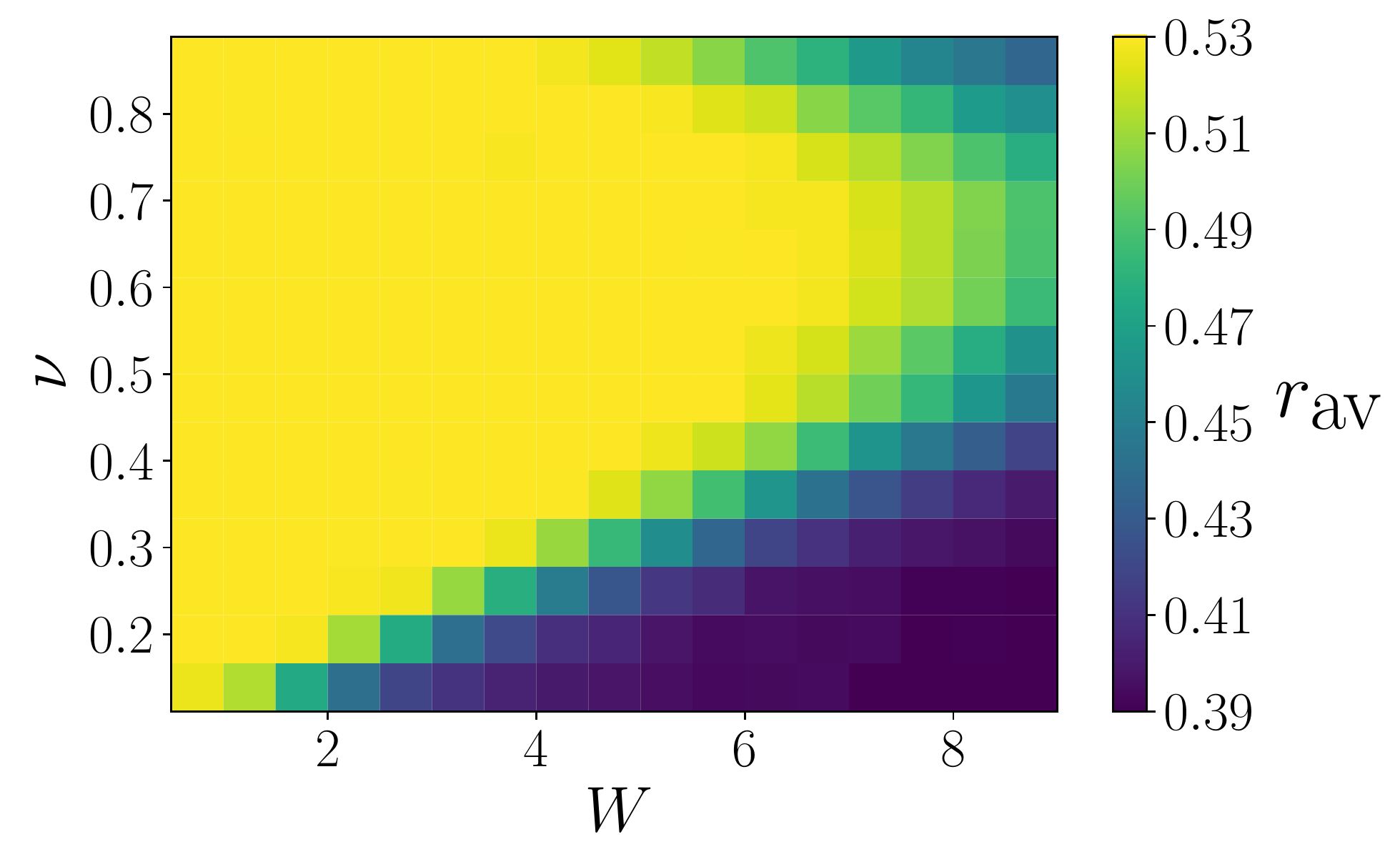}\\
\caption{The sharp difference of $r_\text{av}$ obtained for different $\nu$ at the same disorder $W$ clearly shows the MBME in our model. Interestingly, the mobility edge curve $W_c(\nu)$ is not symmetric, but is peaked around $\nu=2/3$, implying that the states with the maximum number of pairs for fixed size are the hardest to localize. The data is obtained for a system of size $L=18$, using shift-invert method with $10-10^3$ states from the middle of the spectrum and $5\times10^4-10^3$ disorder realizations.\label{Fig:r-ME}}
\end{center}
\end{figure}

While in the main text we focused on the two values of filling, $\nu=1/2$ and $1/5$, here we demonstrate the density dependence of critical disorder. For this purpose we calculate the average ratio of level spacings $r_\text{av}$ for a single system size $L=18$ at varying values of density. Figure~\ref{Fig:r-ME} allows to estimate the dependence of the critical disorder on the filling, $\nu$.  At low densities ($\nu<\nu_c(W)$) states  have $r_\text{av}$ approaching value characteristic for Poisson distribution of level spacings. In contrast, for dense configurations ($\nu>\nu_c(W)$) the level spacing ratio is close to GOE prediction.

Figure~\ref{Fig:r-ME} reveals that the most delocalized filling is $\nu=2/3$, which corresponds to the case when the best packing of pairs in the chain, $\bullet\bullet\circ\bullet\bullet\circ\cdots$,  can be achieved. At this filling the Poisson values of $r_\text{av}$ would be achieved beyond the upper limit of the considered disorder range. Decreasing particle density away from this value causes earlier onset of localization. For instance, fixing disorder value $W=6.5$ we observe that $\nu=1/2$ and $\nu=1/5$ are situated well in delocalized  and localized regions. 

\begin{figure}[t]
\begin{center}
\includegraphics[width=.95\columnwidth]{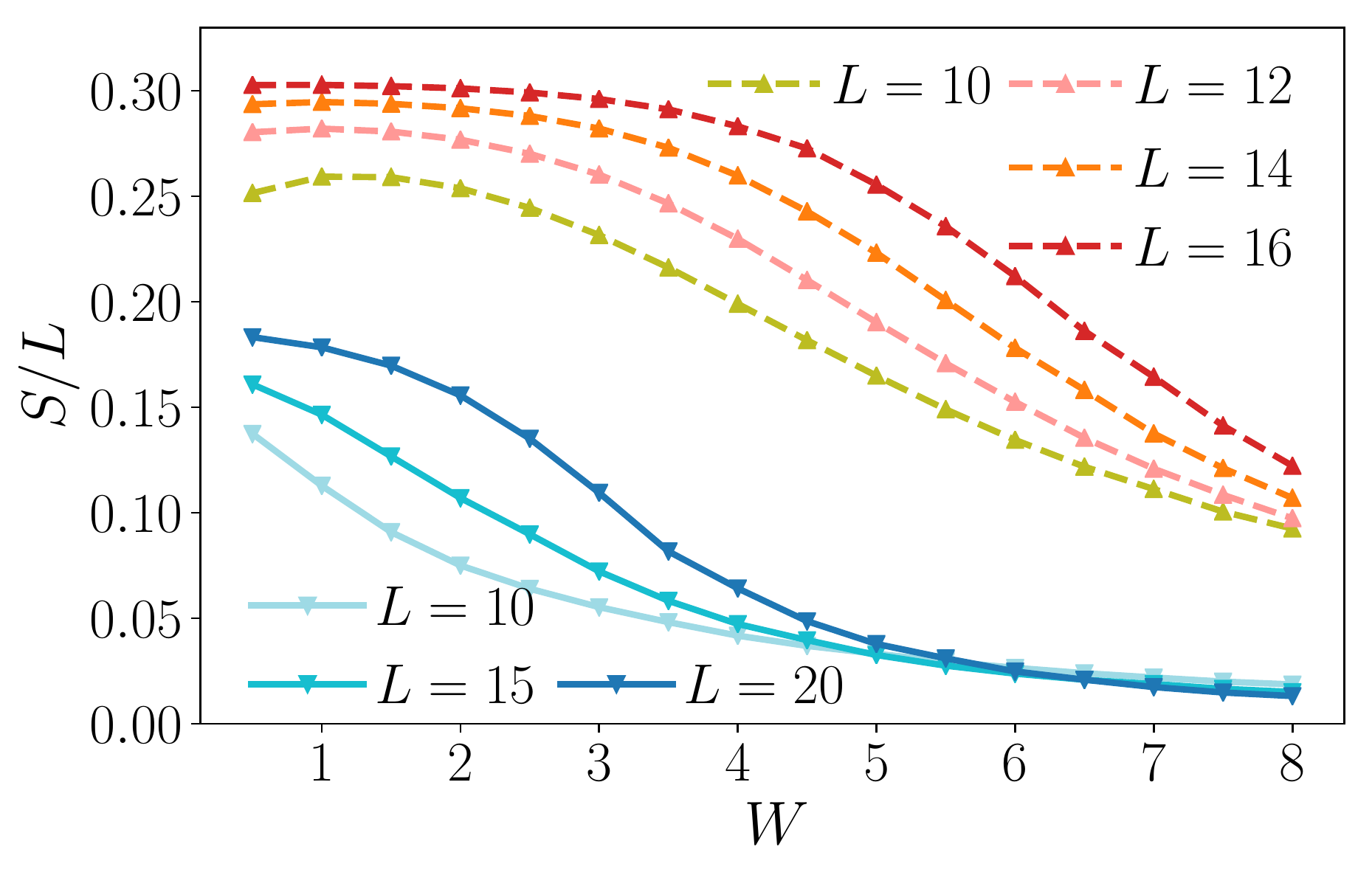}
\caption{ The behavior of half-chain entanglement entropy shows very distinct behavior for dilute (blue-shaded curves) and dense (red-shaded curves) states. The crossing in the dilute states implies that they entered the MBL phase, and thus have area-law entanglement entropy. On the other hand, dense states do not show a similar crossing in this range of disorder, suggesting that they are still in the ergodic phase. The data are obtained with shift-invert method for $10-10^3$ eigenstates in the middle of the spectrum and averaged over $5\times 10^4-5\times 10^3$ disorder realizations. \label{Fig:SED}}
\end{center}
\end{figure}

In addition to the level statistics indicator presented in the main text, we studied other commonly used probes of ergodicity. In particular, Fig.~\ref{Fig:SED} illustrates the behavior of bipartite entanglement entropy for different disorder strengths and different fillings. On the one hand, the finite size scaling of entanglement entropy of eigenstates in the middle of the spectrum shows that for $\nu=1/5$ and disorder $W>W_c\sim 6$ the entanglement is consistent with area-law. On the other hand, the entanglement of dense systems, $\nu=1/2$, does not show a similar behavior. The finite size scaling, indeed, shows no crossing at these disorder values, thus suggesting volume-law of entanglement entropy for $\nu=1/2$.

\subsection{Energy Density MBME}
In this section we briefly discuss the presence of many-body mobility edge in energy density in a single density sector of the Hamiltonian Eq.~(\ref{Eq:H}). Given the $U(1)$ symmetry, we would expect that a fixed filling sector presents MBME in energy density, similarly to the case of the random field XXZ spin chain. In the middle of the spectrum the density of states is large and eigenstates may remain delocalized, while at the same disorder strength the states at the edges of the many-body spectrum are localized.

\begin{figure}[b]
\includegraphics[width=.99\textwidth]{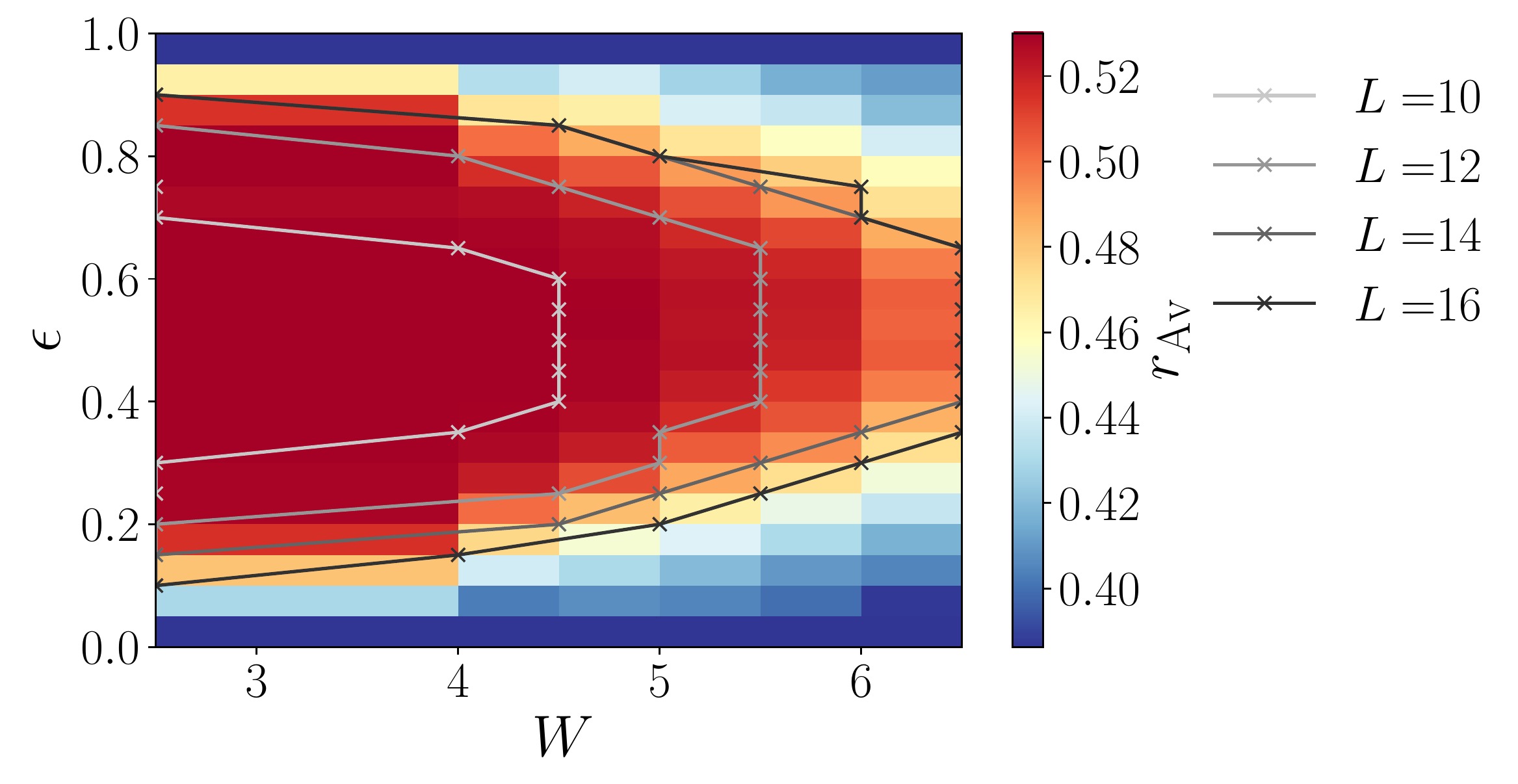}
\caption{ The plot, presenting the energy resolved level spacing ratio for $L=16$ in the half filling sector, shows clear evidence of MBME. In the center of the band $r_\text{av}$ approaches the GOE value, while at large and small energy density it is close to the Poisson value. Furthermore, the MBME curves obtained for smaller systems seem to converge at increasing system size, thus suggesting the persistence of the MBME in the thermodynamic limit.
The plot was obtained averaging over $n=10^4,\;2\times10^3,\;5\times10^2,\;10^2$ disorder realizations for increasing system size from $L=10$ to $L=16$.
\label{Fig:energy mbmbe}}
\end{figure}
\begin{figure*}[t]
\begin{center}
\includegraphics[width = .45\textwidth]{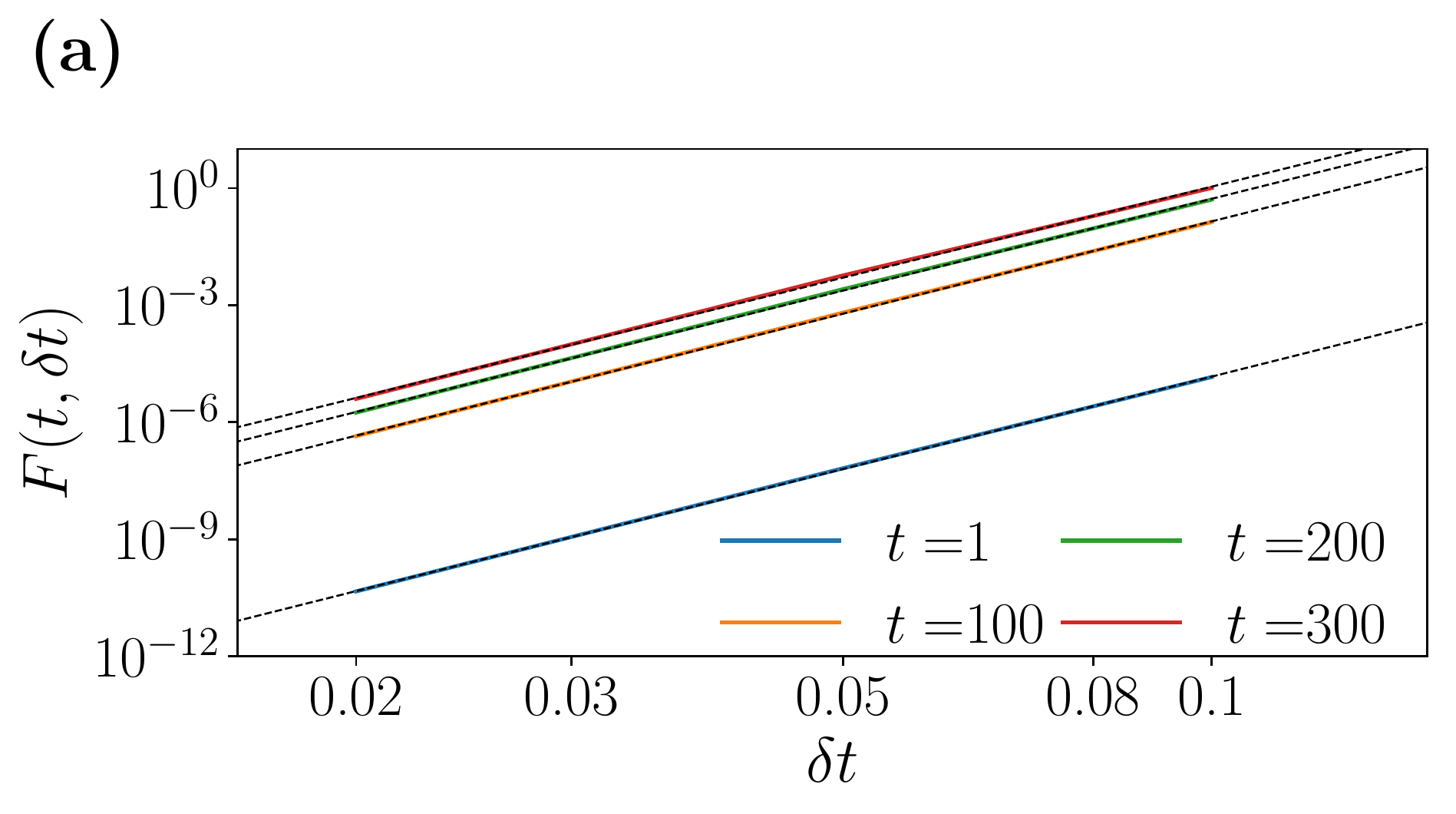} \quad \includegraphics[width = .45\textwidth]{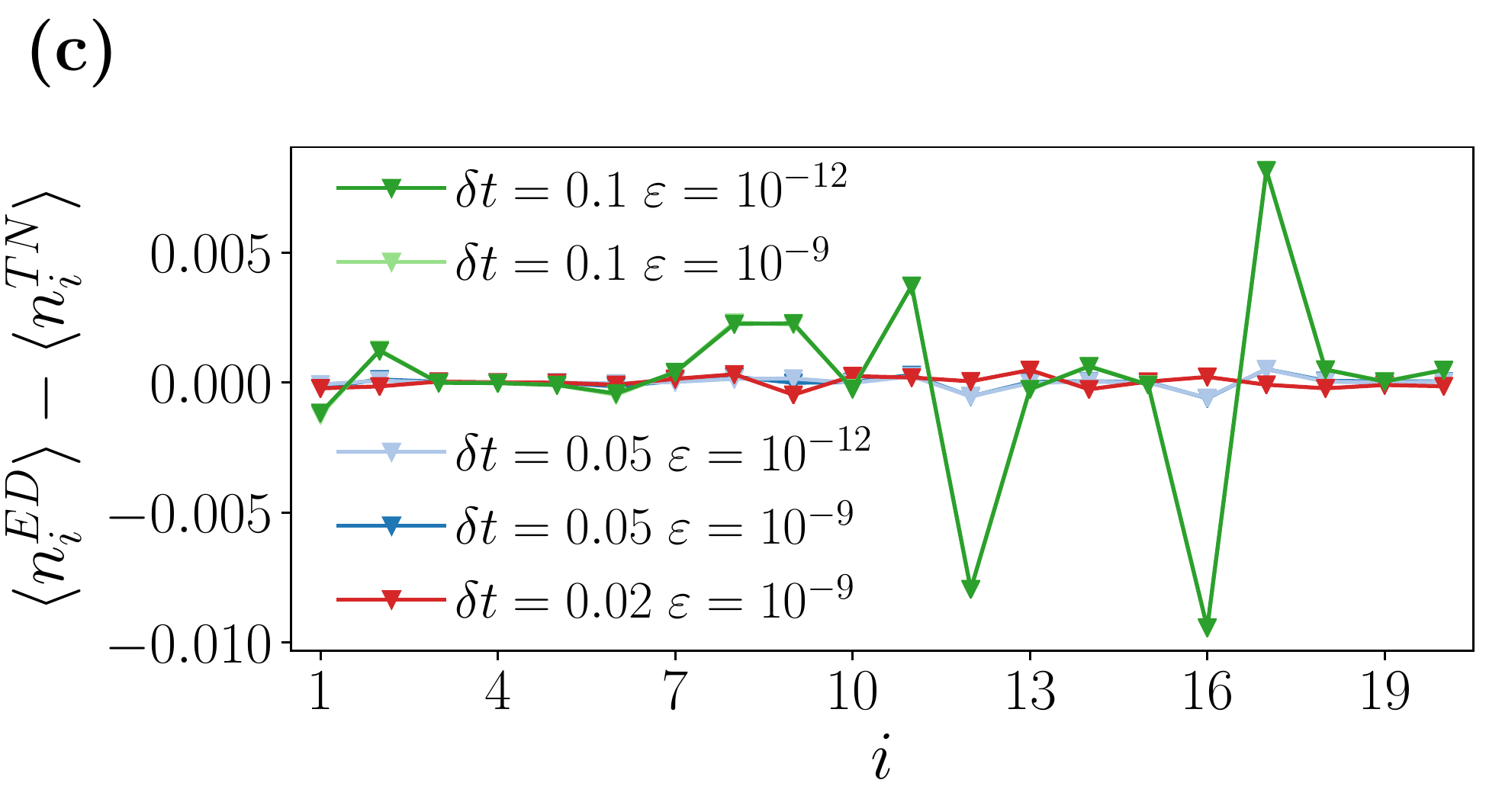} \\
\includegraphics[width = .45\textwidth]{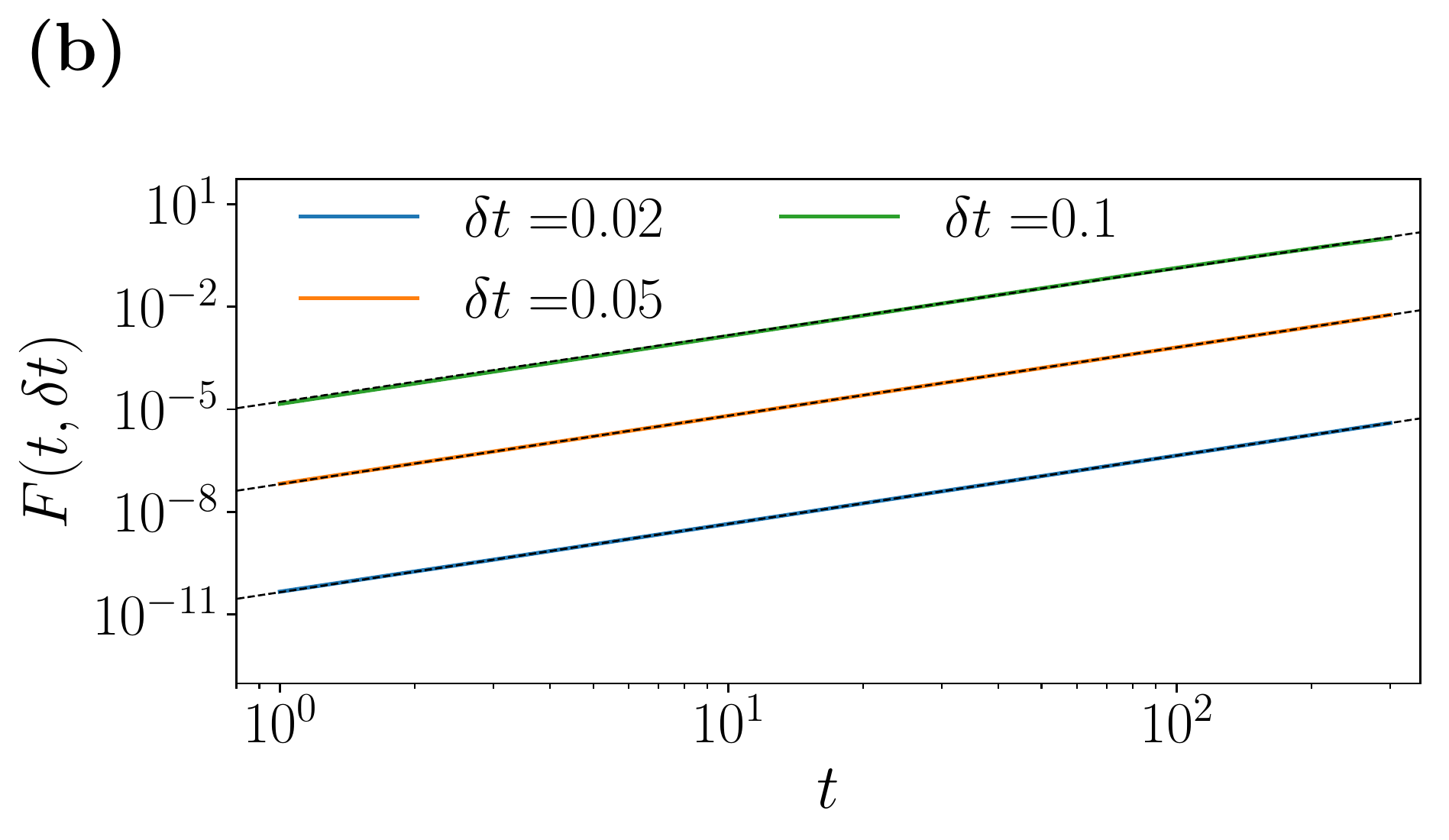} \quad \includegraphics[width = .45\textwidth]{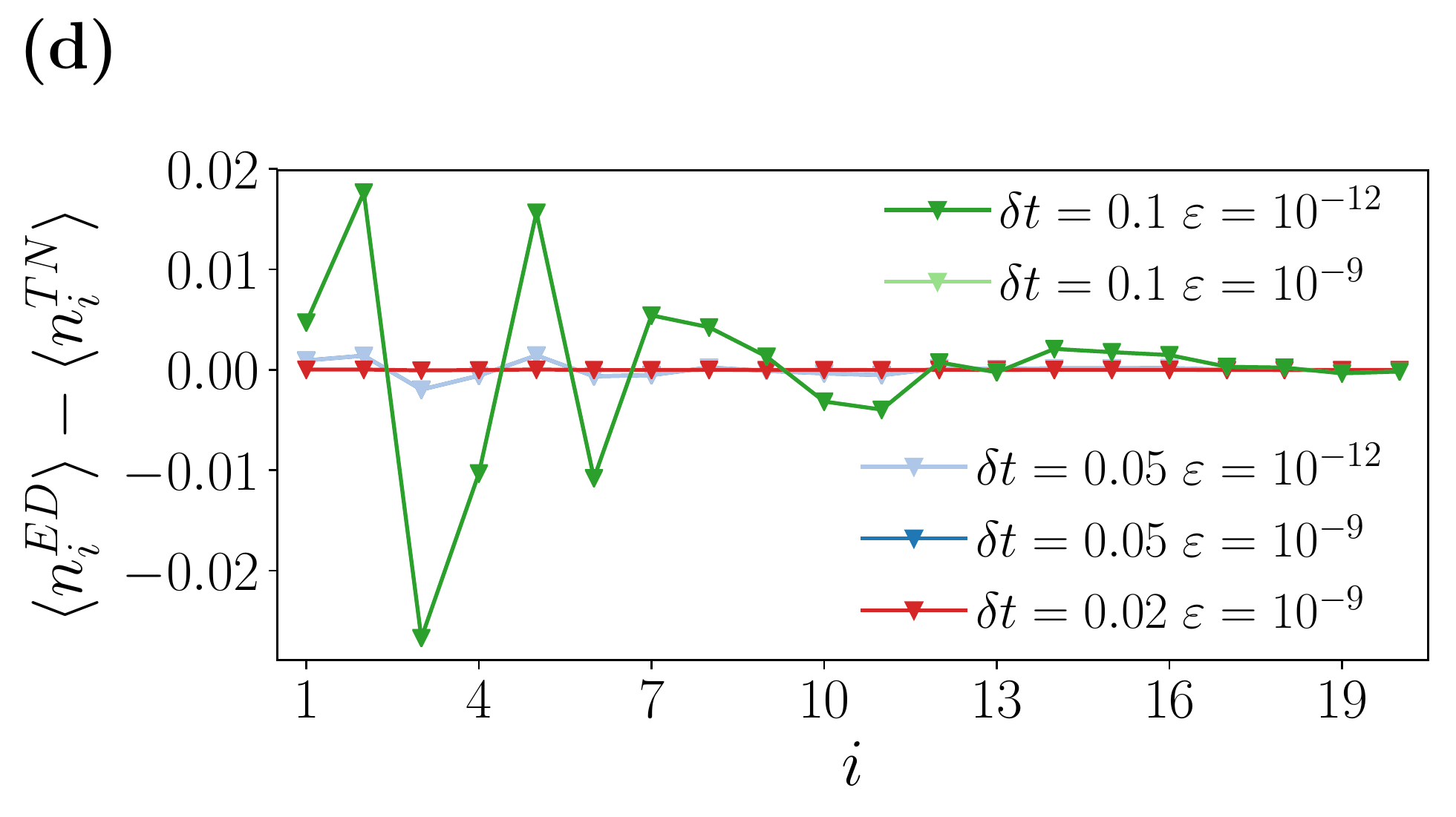} 
\caption{(a-b) Deviation of ground state fidelity from 1 in Trotter time evolution, $F(t,\delta t)$, shows power-law behavior both in time and in time-step, as expected. The data is obtained at density $\nu = 1/5$, system size $L=20$ and disorder $W=6.5$ for a particular disorder realization. The plots for other disorder realizations are qualitatively similar. (c-d) The comparison between ED and TEBD time evolution reveals that the most effective way to increase accuracy of TEBD is to decrease the time-step $\delta t$. Indeed, the change in the truncation between $\varepsilon = 10^{-9}$ and $\varepsilon = 10^{-12}$  does not have much effect on the the difference between density profiles of exact diagonalization and TEBD. At the same time, the decrease of time step brings the local density profile closer to ED results.  The density profiles are calculated by propagating uniform density wave (c) and uniform pair-density wave (d) initial states to time $t=500$ for a particular disorder realization with $L=20$, $\nu=1/5$, $W=6.5$. \label{Fig:MPS}}
\end{center}
\end{figure*}

To explore the eventual presence of MBME in energy density in the half filling sector, we studied the energy resolved level spacing ratio $r_\text{Av}(\epsilon,W)$. The results, displayed in figure~\ref{Fig:energy mbmbe}, show evidence of many-body mobility edge; the level spacing ratio, as a function of the energy density $\epsilon = \frac{E-E_\text{min}}{E_\text{max}-E_\text{min}}$, where $E_\text{min}$ and $E_\text{max}$ are the ground state and the most excited state respectively, increases from the Poisson to the GOE value as $\epsilon$ goes from the lower edge to the center of the spectrum and decreases again from the center to the upper edge. This variation is such that at a fixed disorder strength the low and high energy states are localized, while the center of the band is delocalized, thus defining a many-body mobility edge. The scaling of the MBME curves for different system size shows signs of convergence, suggesting stability of the MBME in the thermodynamic limit.

\begin{figure}[t]
\begin{center}
\includegraphics[width=0.95\columnwidth]{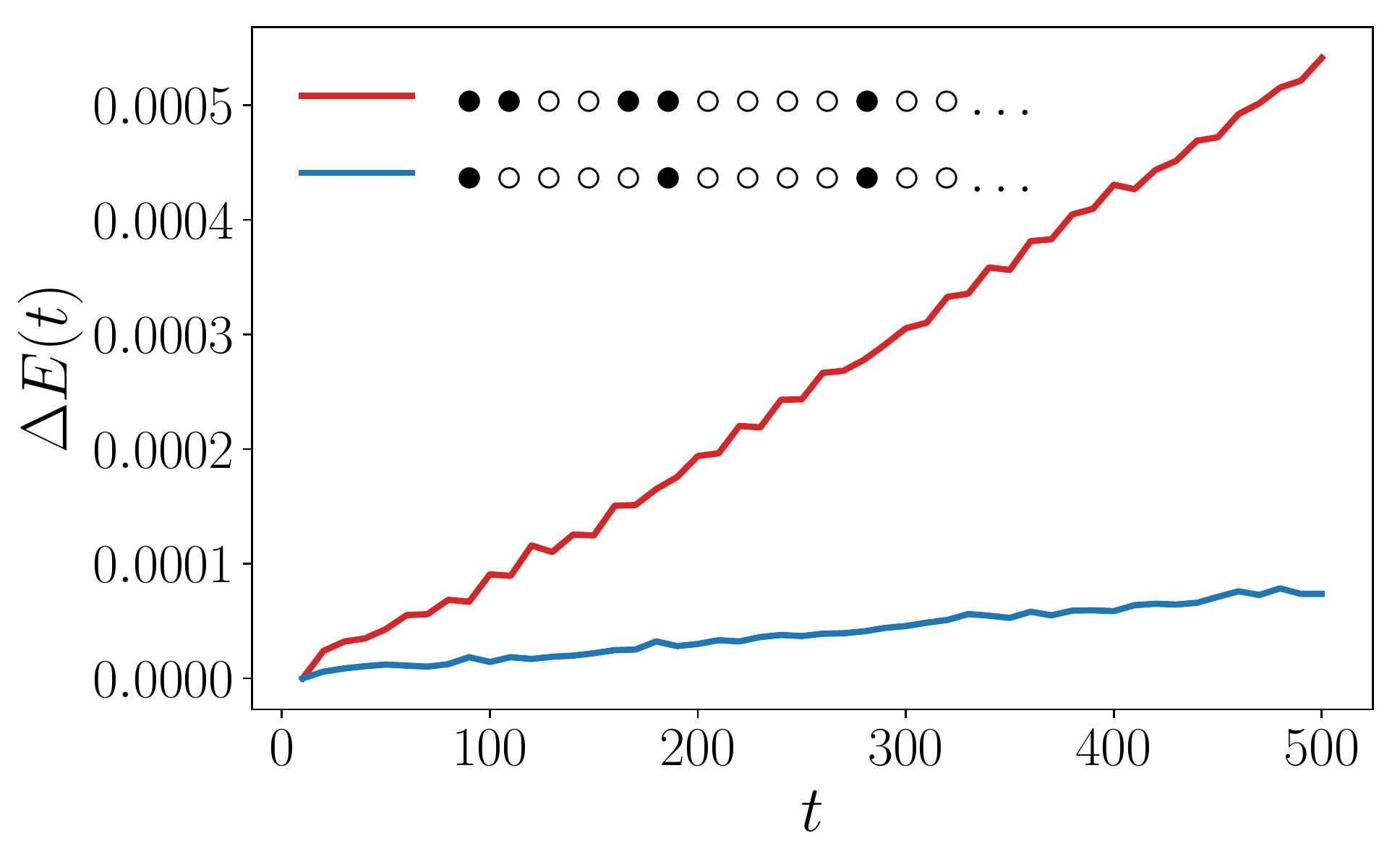}
\caption{The normalized absolute value of the energy difference from the initial energy $\Delta E(t)=|\langle H(t)\rangle -E(0)|/|E(0)|$ remains very small for both the density wave, blue curve, and the non-uniform, red curve, configurations, confirming the good accuracy of our numerical simulations beyond the ED benchmark. The larger deviation displayed by the non-uniform configuration is understood as a result of the presence of the bubble in the lattice, that increases entanglement growth.
The results here shown are obtained averaging over $100$ disorder realizations, for the system sizes, $L=30$, and initial states described in the main text. 
\label{Fig:DeltaE}}
\end{center}
\end{figure}

\section{MPS simulations of quench dynamics}\label{Sec:MPS}
In our MPS simulation, we time evolve dilute states in large systems $L\geq 30$ up to time $T_\text{max} = 500$. For this we use the time-evolving block decimation (TEBD) algorithm with a fourth-order Trotter evolution based on the ITensor library~\cite{ITensor}. The main parameter involved in the time evolution algorithm is the time step $\delta t$ used to split the unitary evolution into a sequence of gates. The error related to the finite size of the time step in the $p$-th order Trotter expansion grows as $\delta t^p$. The other source of error is the finite cutoff, $\varepsilon$, that governs the truncation of singular values in the singular value decomposition (SVD).

While the instantaneous errors related to the truncation and finite time step are known, understanding the propagation of these errors with time and their possible interference is challenging. First we tested TEBD algorithm by evolving the ground state of the same model. Provided that the time evolution is numerically exact, the overlap between the TEBD-evolved ground state, $\ket{\psi_0(t)}=U^\text{TEBD}(t)\ket{GS}$ and the exact time evolution of the ground state, $\ket{GS(t)}=e^{-\imath E_0 t}\ket{GS}$, is supposed to give the identity $\bra{\psi_0(t)}\ket{GS(t)}=1$ at all times. For the fourth-order Trotterization the behavior of $F = 1-|\bra{\psi_0(t)}\ket{GS(t)}|$  is known to be proportional to $(\delta t)^8$. The numerical results plotted in Fig.~\ref{Fig:MPS}(a), confirm these expectations. 

Next, we performed a benchmarking of TEBD algorithm against ED time evolution for several disorder realizations and simulation parameters. An illustration of such benchmarking is shown in Fig.~\ref{Fig:MPS}(c) and (d). In particular, we observed that time step $\delta t =0.05$ and cutoff $\varepsilon=10^{-9}$ result in a good agreement between ED and TEBD dynamics. Smaller values of $\delta t,\varepsilon$ would improve the agreement but would result in a dramatic slowdown of the evolution time. Therefore, we decided to use these parameters in the simulations presented in the main text.

When larger system sizes are involved, as it is the case for the simulations actually used in the main text, comparison with exact results is not available. Therefore, other indicators for the accuracy must be studied. Among these, energy conservation through the time evolution is a straightforward probe. The energy deviation $\Delta E(t) = |E(0)-E(t)|/E(0)$, where $E(t)=\bra{\psi(t)}\hat{H}\ket{\psi(t)}$, allows to control the propagation of the error during the Trotter time evolution. In Fig.~\ref{Fig:DeltaE} we show the results for $\Delta E(t)$ in the two quenches presented in the main text, for $L=30$. The two plots highlight that the average energy deviation is very small in both configurations. In spite of that, a clear difference can be observed among the two quenches, noticing that the non-uniform state has larger error. This is probably due to the enhanced entanglement caused by the presence of the bubble in the lattice. Nevertheless, $\Delta E(t)$ remains very small even at long times, thus confirming the reliability of our long-time numerical simulations.

\begin{figure}[b]
\begin{center}
\includegraphics[width = .45\columnwidth]{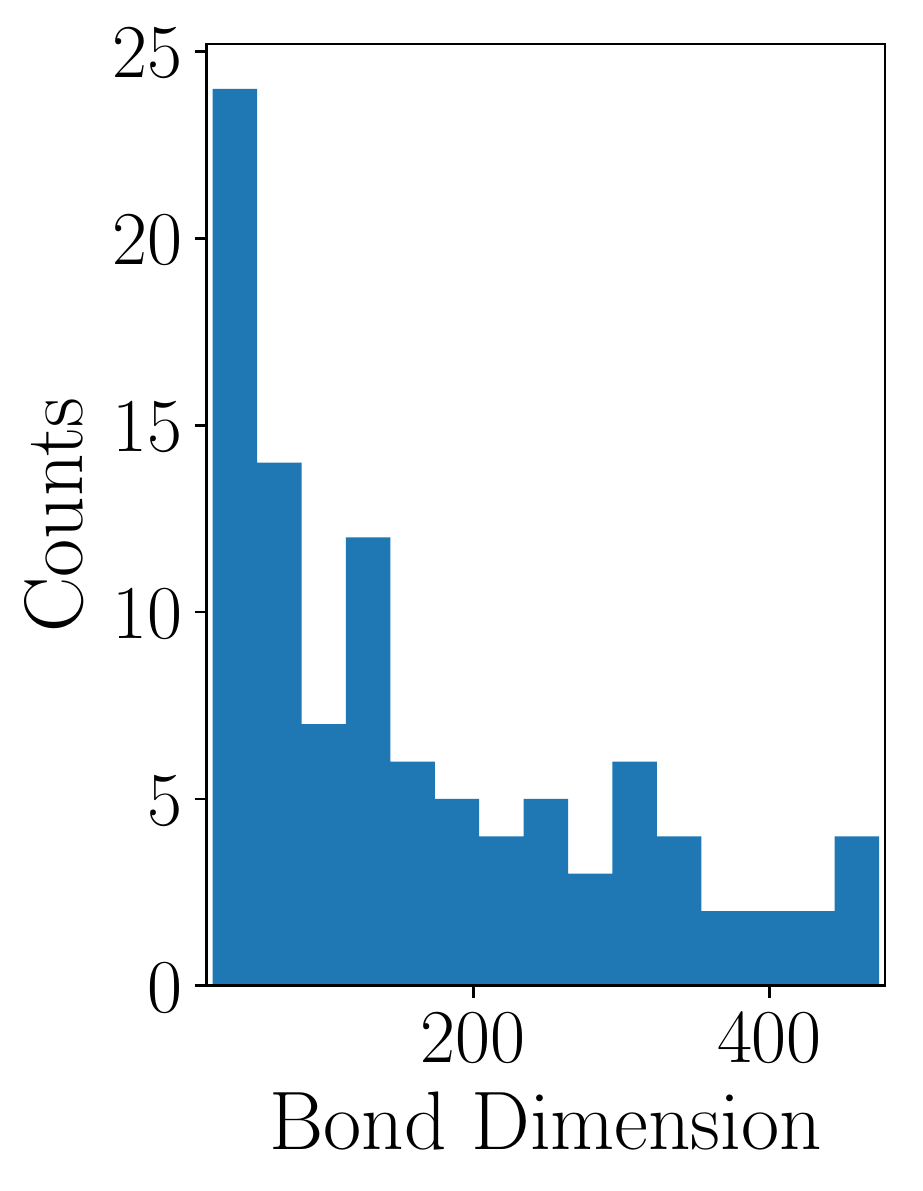} \quad
\includegraphics[width = .45\columnwidth]{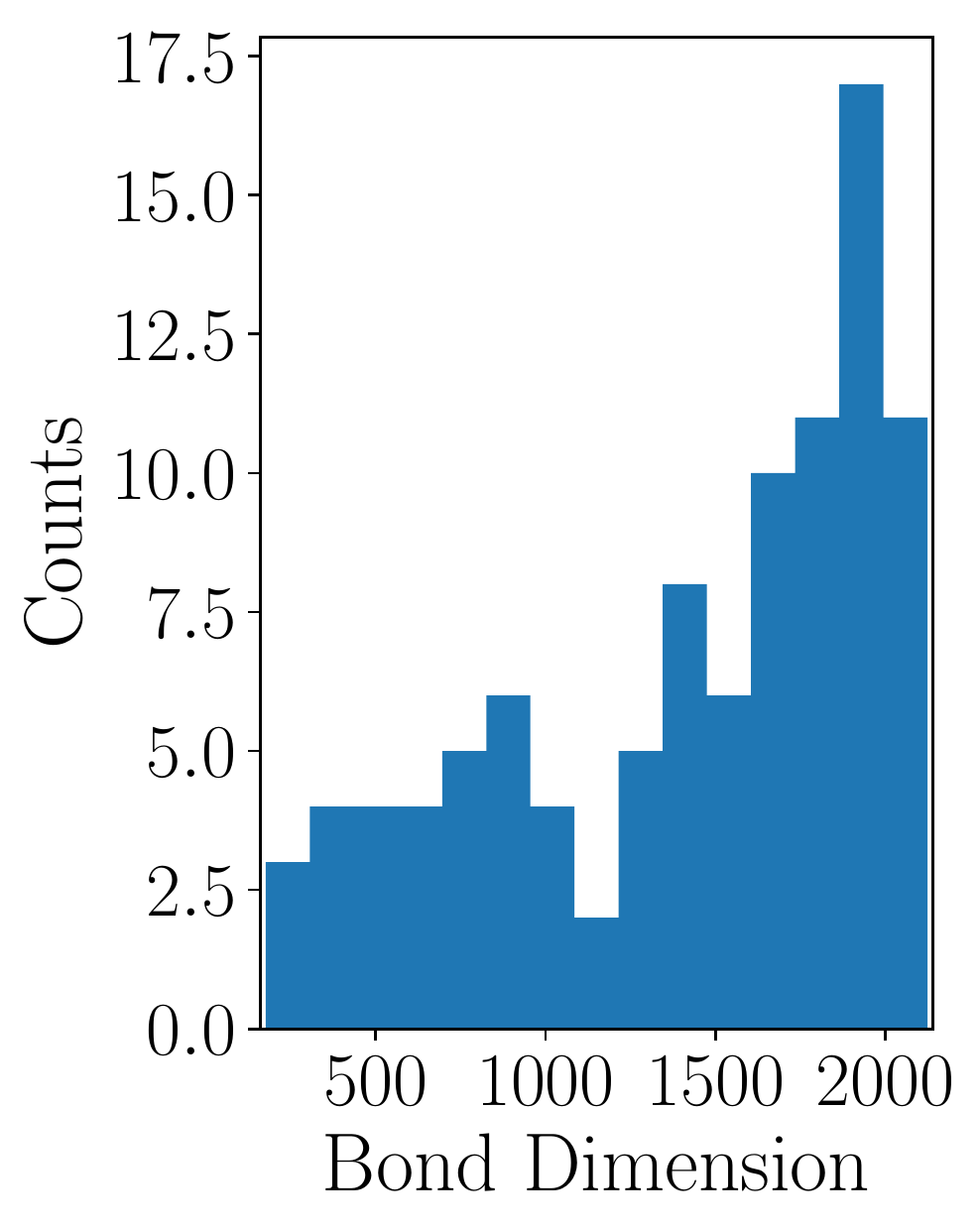} 
\caption{The histogram representing the maximum bond dimension of different disorder realizations show that the threshold values of $500$ and $3000$ for the uniform density wave (left) and bubble states (right) were never saturated in our simulations.\label{Fig:bond}}
\end{center}
\end{figure}

In all the simulations performed using ITensor~\cite{ITensor}, we used the $U(1)$ symmetry implementation. In particular, to obtain the numerical results presented in Fig.~\ref{Fig:quench}(a) and (d) we set the maximum bond dimension to be $500$ and $3000$ respectively. As the histograms in Fig.~\ref{Fig:bond} show, all the disorder realizations remained well below the maximum threshold. This fact ensures that we have a control on the error encountered in the evolution, in contrast to time evolution with TDVP with fixed bond dimension, where error estimation is more challenging~\cite{Doggen18}.

\section{Additional results from quench dynamics}

In the main text we discussed dynamics in quenches that begin from a uniform state or a bubble joined to a more dilute remainder. Below we present details for the quenches in presence of bubble.  In addition, we discuss the dynamics resulting from the initial state containing density wave of particle pairs.
\subsection{Pair density and entanglement dynamics in presence of a bubble}
Since particle pairs are the most mobile objects, we consider the pair density in quenches that are initialized with the bubble (see Fig.~\ref{Fig:quench}(d),(g) in the main text). The pair density is of special interest in these quenches as in Ref.~\cite{DeRoeck2016} suggested that the instability of the system is ascribed to the ability of the bubble to move. In our model the bubble consists of several pairs, thus motion of the bubble throughout the system would imply the spreading of pairs.

\begin{figure}
\begin{center}
\includegraphics[width=.95\columnwidth]{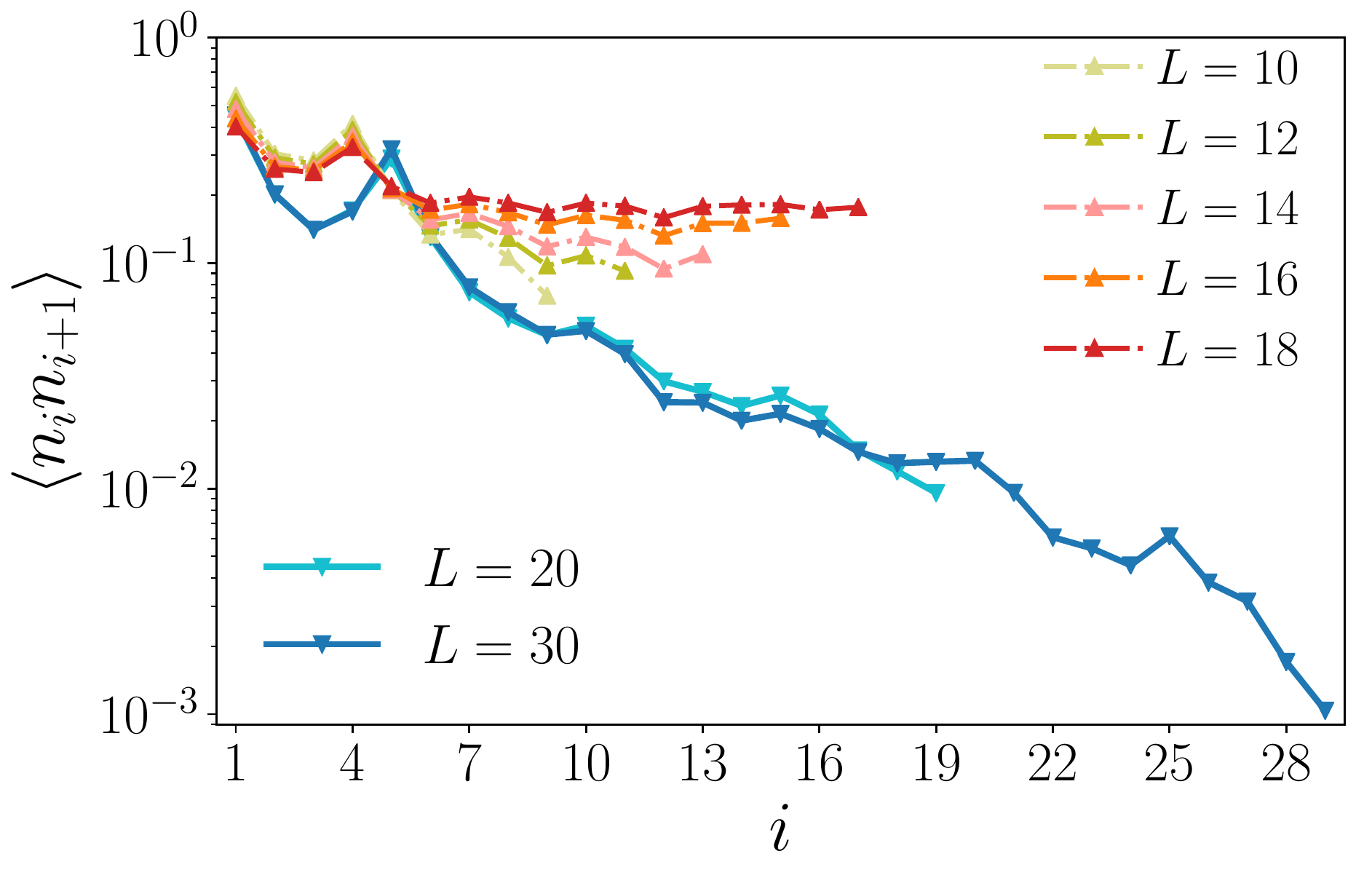}
\caption{ The finite size scaling of the pair density $\langle n_i n_{i+1}\rangle$ shows opposite trend for the dense and dilute cases.
The red-shaded curves represent $\nu=1/2$ configurations: increasing the system size (from yellow to dark red) the pair density becomes more uniform and approaches the thermal value, hence in the thermodynamic limit the probability of finding a pair far from the bubble is almost the same as finding it in the bubble. On the contrary, blue curves ($\nu=1/5$) show exponential vanishing of the pair density and, furhtermore, increasing system size (from light blue to dark blue) the density decreases, suggesting that at the thermodynamic limit there will be no pair outside the thermal region.
Data were obtained with ED, Krylov ($T_\text{max}=1000$) and TEBD ($T_\text{max}=500$) algorithms averaging over $100$ disorder samples for the largest MPS simulations ($L=20,30$), $3\times10^4,10^4,5\times10^3\text{ and }10^3$ for ED (from $L=10$ to $L=16$) and over $10^3$ for Krylov algorithm ($L=18$). 
\label{Fig:nP}}
\end{center}
\end{figure}

The pair density defined as $\langle n_i n_{i+1}\rangle$ measured at late or infinite times is shown in Fig.~\ref{Fig:nP}.  In the dense case the late time pair density profile supports delocalization: at late times the density of pairs becomes homogeneous throughout the formerly more dilute region of the system. We note, that the pair density is not a conserved quantity, and it can increase in the process of unitary dynamics. 

In contrast, for the dilute case the pair density profile has a pronounced exponential tail away from the initial ergodic region. This shows that pairs spreading away from the initial bubble do not delocalize when encountering additional particles on their way. Indeed, while the late time pair density profile has small peaks around the initial position of particles, these peaks are not very pronounced. In addition, the study of the  pair density profile in the uniform density wave at $\nu=1/5$ reveals an almost constant behavior, centered around $\langle n_i n_{i+1}\rangle\sim 10^{-3}$, which corresponds to the values reached at the end of the exponential tail in the system with $L=30$ in Fig.~\ref{Fig:nP}.

Next, we focus on understanding different contributions to entanglement growth. Exploiting the $U(1)$ symmetry of our model and following Refs.~\cite{islam2015,lukin2018}, we split the von Neumann entanglement entropy into a configuration and a particle transport contributions. Indeed, due to conservation of the total number of bosons the full reduced density matrix $\rho$ must have a block-diagonal form. Individual blocks within $\rho$ can be written as $p_n \rho^{(n)}$, where $p_n$ gives the probability to have $n$ particles in the susbsystem $A$  and $\rho^{(n)}$ is normalized as $\tr \rho^{(n)}=1$.  Using such representation of the reduced density matrix we can split the full entropy into $S_{vN}=S_C+S_n$ as:
\begin{equation}
    \label{Eq:SvN=Sn+Sc}
    \begin{split}
        S_{vN} &= -\tr \rho\log \rho = -\sum_{n}p_n\tr \rho^{(n)} \log(p_n\rho^{(n)}) \\
        &= -\sum_n p_n\log p_n - \sum_{n}p_n\tr \rho^{(n)} \log \rho^{(n)}\\
        &=S_n+S_C.
    \end{split}
\end{equation}

\begin{figure}[t]
\begin{center}
\includegraphics[width=.95\columnwidth]{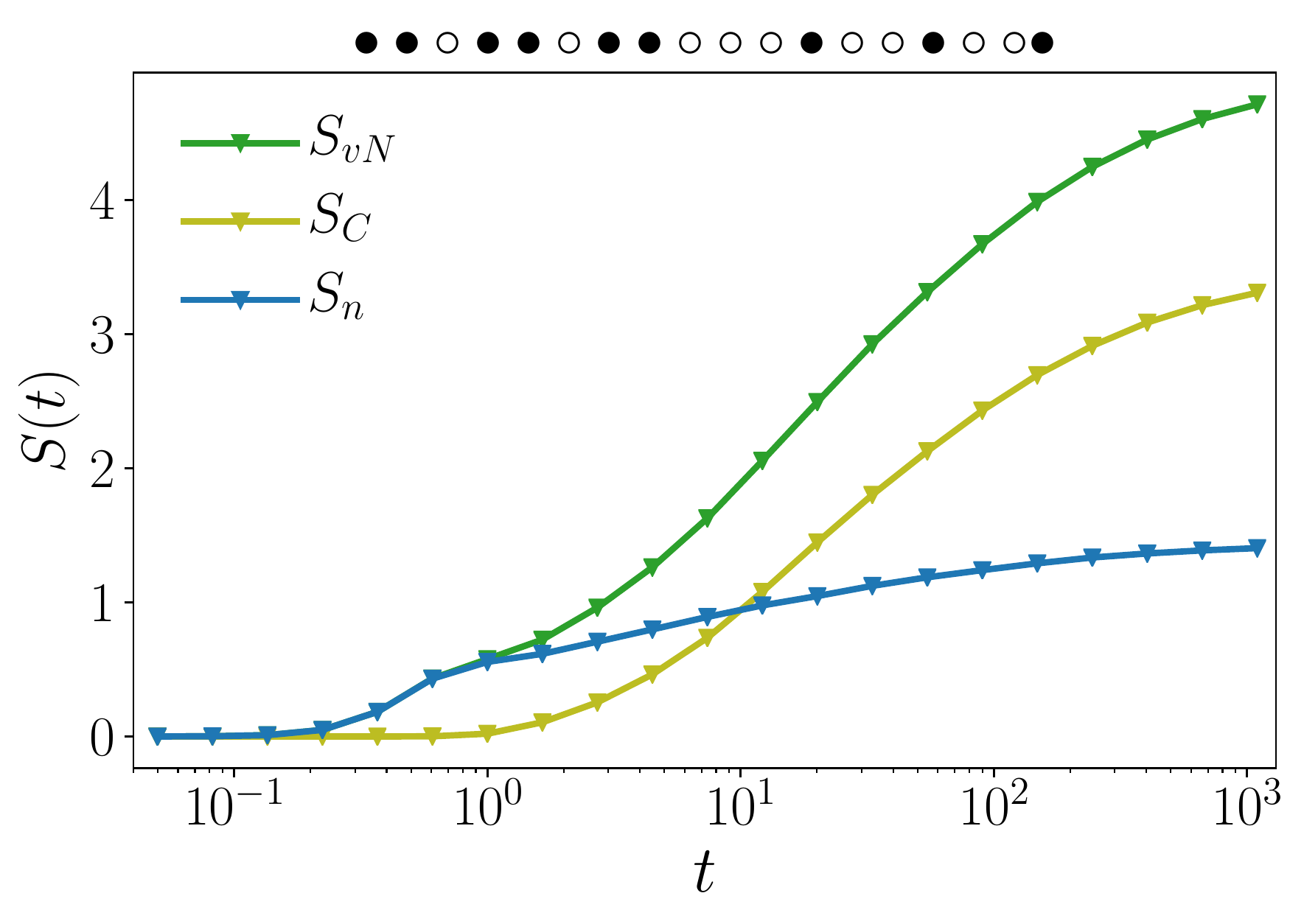}
\caption{The different contributions to the entanglement entropy of the bubble show that the overall behavior of the von Neumann entropy is faster than logarithmic. Nevertheless this behavior can be ascribed to the sole configurational entropy $S_C$, while the particle transport contributes to the purely logarithmic growth.
The curves are obtained through Krylov evolution up to $T_\text{max}=1000$ averaged over $1000$ disorder realizations for $L=18$ and $W=6.5$.
\label{Fig:Quench S}}
\end{center}
\end{figure}

\begin{figure}[hb]
\includegraphics[width=.9\textwidth]{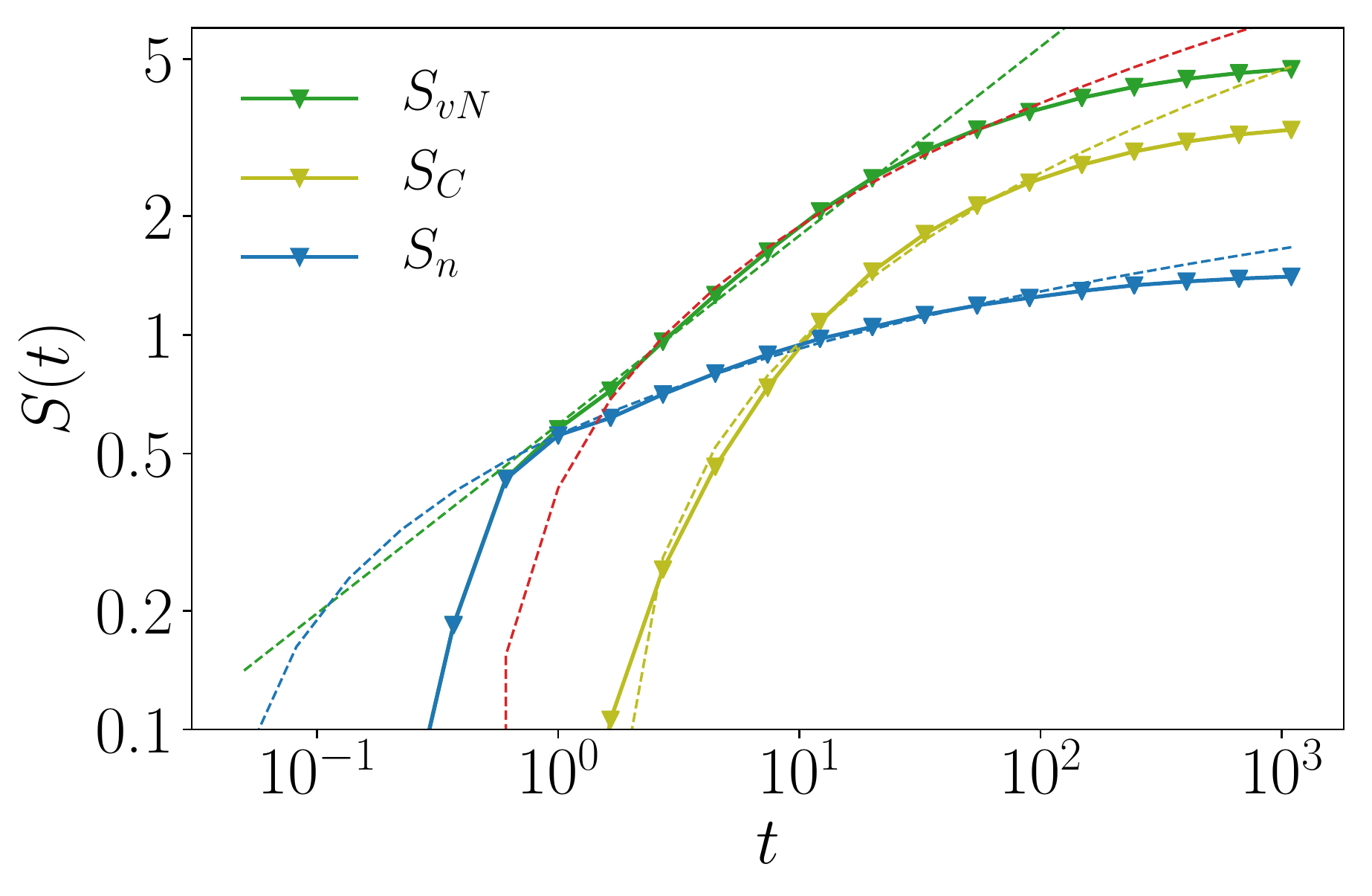}
\caption{The log-log plot of the entropies discussed in Eq.~(\ref{Eq:SvN=Sn+Sc}) confirms that only the total entropy is growing as a power-law, while both $S_n$ and $S_c$ grow slower. In particular, $S_n$ grows as a first degree polynomial in $\log(t)$ (dashed green line) and $S_C$ as a second degree polynomial in $\log(t)$ (orange dashed line). The sum of these two behaviors (red dashed line) agrees with the power-law behavior (dashed blue line).
\label{Fig:S-fits}}
\end{figure}

In this way the entanglement growth is split into two contributions: one coming from the particle transport, and another originating from dephasing between different configurations with the same particle number. Interestingly, Fig.~\ref{Fig:Quench S} shows that while the overall entanglement entropy grows faster than logarithmic, this is due only to the configuration part (yellow curve) and the entanglement due to particle transport has logarithmic growth. The logarithmic growth of $S_n$ is consistent with the logarithmic particle transport presented in Fig.~\ref{Fig:quench}(h) in the main text and with other transport measures presented in the next section. We identify this behavior as a hallmark of MBME, and note that it happens on long, yet experimentally accessible timescales $t\sim 50 (\hbar/t_1)$.

Recent work~\cite{Kiefer-Emmanoulidis2020} demonstrated that the logarithmic growth of the number entropy is expected in the thermal phase, provided there is particle transport over distances that increase as a power-law in time, $l\propto t^\nu$.~\footnote{We note, that although the authors of Ref.~\cite{Kiefer-Emmanoulidis2020}  report unbounded growth of the number entanglement in the MBL phase, the successive work of Luitz and Bar Lev~\cite{Bar Lev 2020} shows that this is due to rare particle fluctuations around the boundary between the two subsystems and the growth disappears at large enough system size.} The authors also predict a power-law scaling of the configuration entropy, which we do not observe, as shown in figure~\ref{Fig:S-fits}. We attribute the slower than power-law growth of configuration entanglement to the localized nature of the right half of the chain, which in turn reduces the number of possible configurations until the particle transport from the left half leads to delocalization. 

Further analysis of the entanglement dynamics shows that $S_{vN}$ grows in a power-law fashion $S_{vN}\approx at^b$, corresponding to the dashed blue fit in Fig.~\ref{Fig:S-fits}, over a relevant time interval. On the other hand, both $S_n$ and $S_c$ behave as polynomials in $\log(t)$, of first and second degree respectively, and their sum (dashed red line) reasonably approximates the power-law behavior of $S_{vN}$, as one can expand $t^b \approx 1 +b\log(t)+\frac{b^2}{2}\log^2(t)$.

\begin{figure}[t]
\quad\quad \includegraphics[width=.8\textwidth]{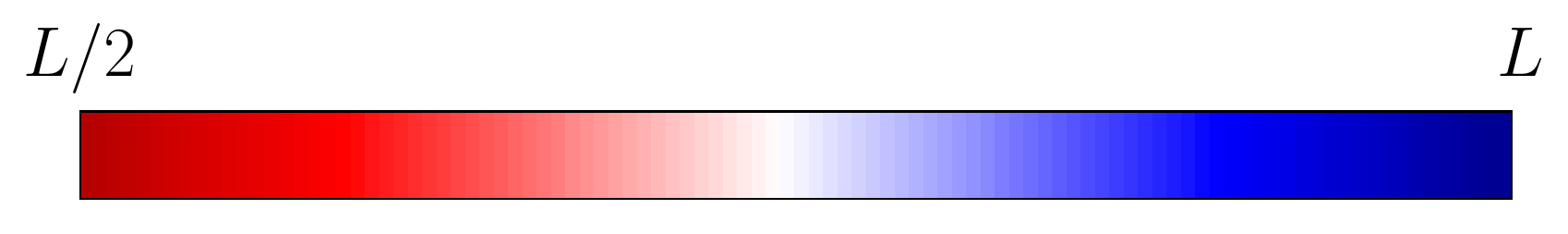}\\
\includegraphics[width=.9\textwidth]{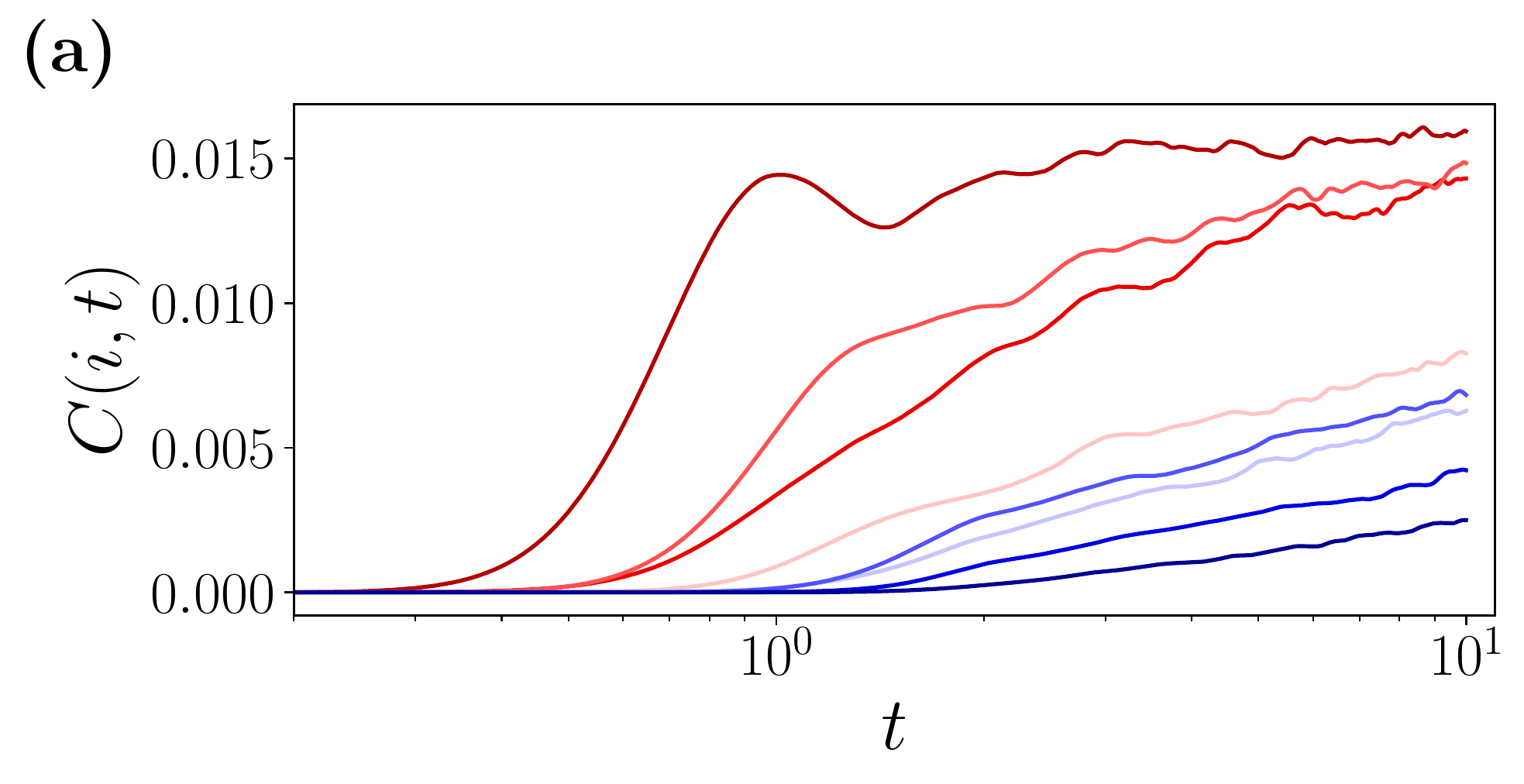}\\ \includegraphics[width=.9\textwidth]{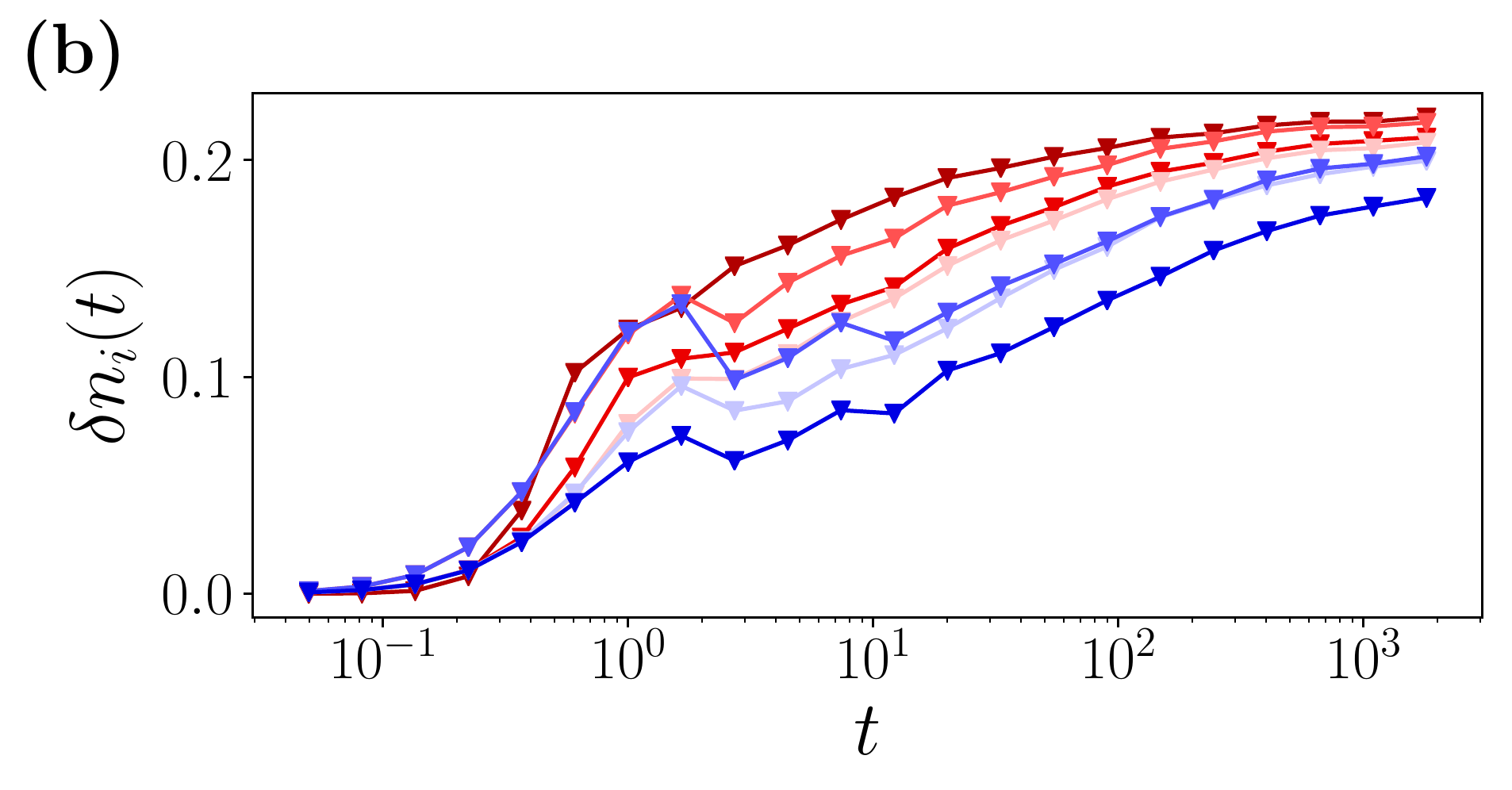}\\
\includegraphics[width=.9\textwidth]{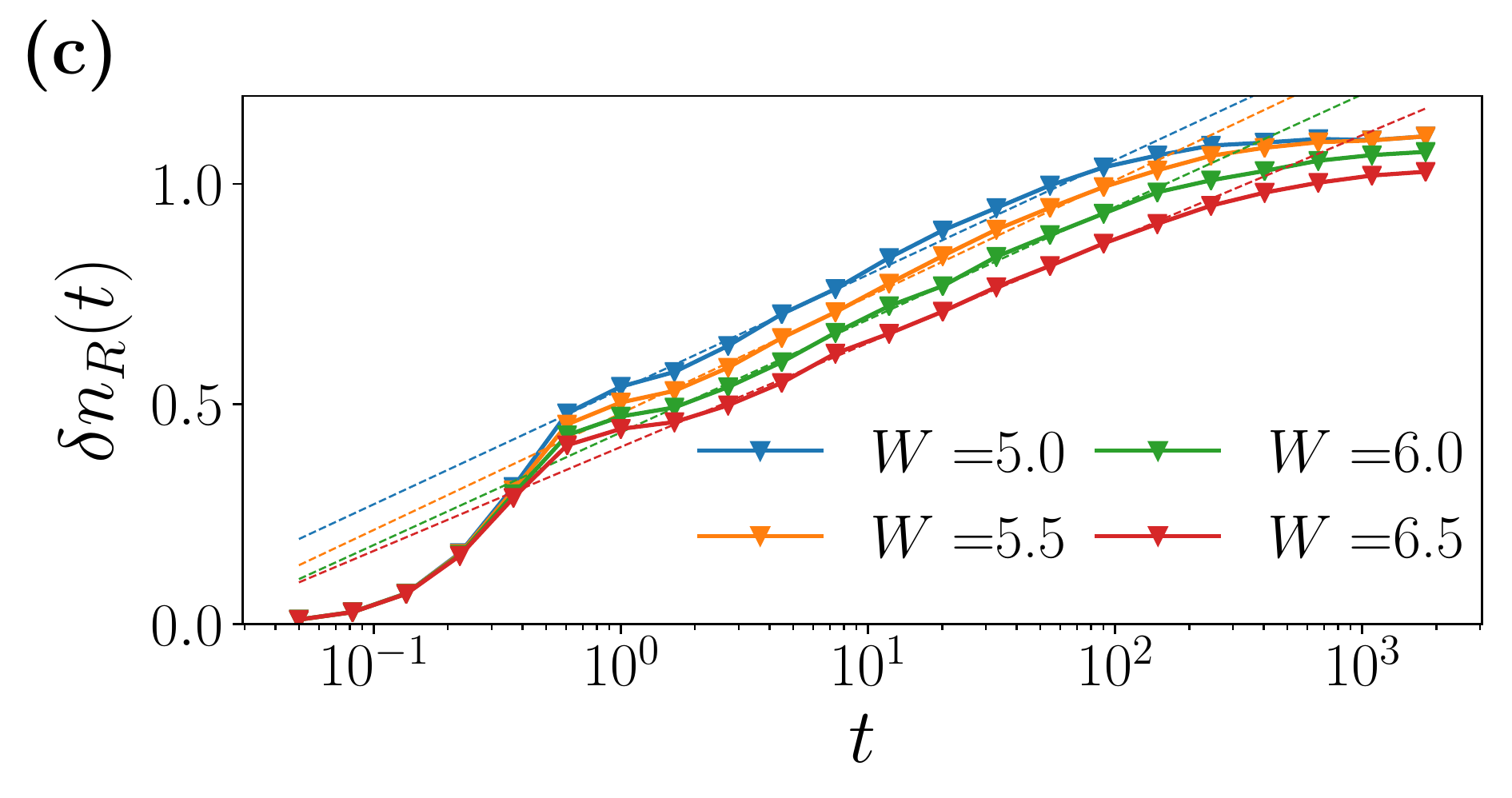}
\caption{Time dynamics of correlation functions~(a), local density fluctuations~(b) and density fluctuations in the dilute half of the chain~(c) all show, after an initial power-law growth, logarithmic increase with time. In particular, correlations far away from the central site (blue curves, as encoded in the legend above) show signs of a logarithmic light-cone. Similarly, local density fluctuations deep in the localized region present slower dynamics. Finally, panel~(c) shows how increasing disorder slows the growth of the global density fluctuation of the dilute half. These results were obtained with the Krylov method on system size $L=18$ for $W=6.5$~(a), with ED on system size $L=16$ and $W=6.5$~(b) and for different disorder values~(c), averaging over $200$ disorder realizations.
\label{Fig:Corr-log}}
\end{figure}

Finally, to support our interpretation of logarithmic increase of $S_n$ as due to transport, we study the dynamics of density correlation functions and fluctuations.  Figure~\ref{Fig:Corr-log} presents the dynamics of connected correlation functions $C(i,t) = \langle \hat{n}_i\hat{n}_{L/2}\rangle-\langle\hat{n_i}\rangle\langle\hat{n}_{L/2}\rangle$ with respect to the central site of the chain, the local density fluctuations $\delta n_i = \langle \hat{n}^2_i\rangle - \langle\hat{n}_i\rangle^2$ and the density fluctuations in the dilute part of the chain $\delta n_R = \langle \hat{n}^2_R\rangle - \langle\hat{n}_R\rangle^2$, where $\hat{n}_R=\sum_{i=L/2}^L\hat{n}_i$. The logarithmic dynamics of these quantities is consistent with the behavior of number entropy, thus proving the further support for the existence of slow transport in the dilute part of the chain.

\subsection{Quench dynamics from a pair density wave state}
\begin{figure}[ht]
\begin{center}
\includegraphics[width=.9\columnwidth]{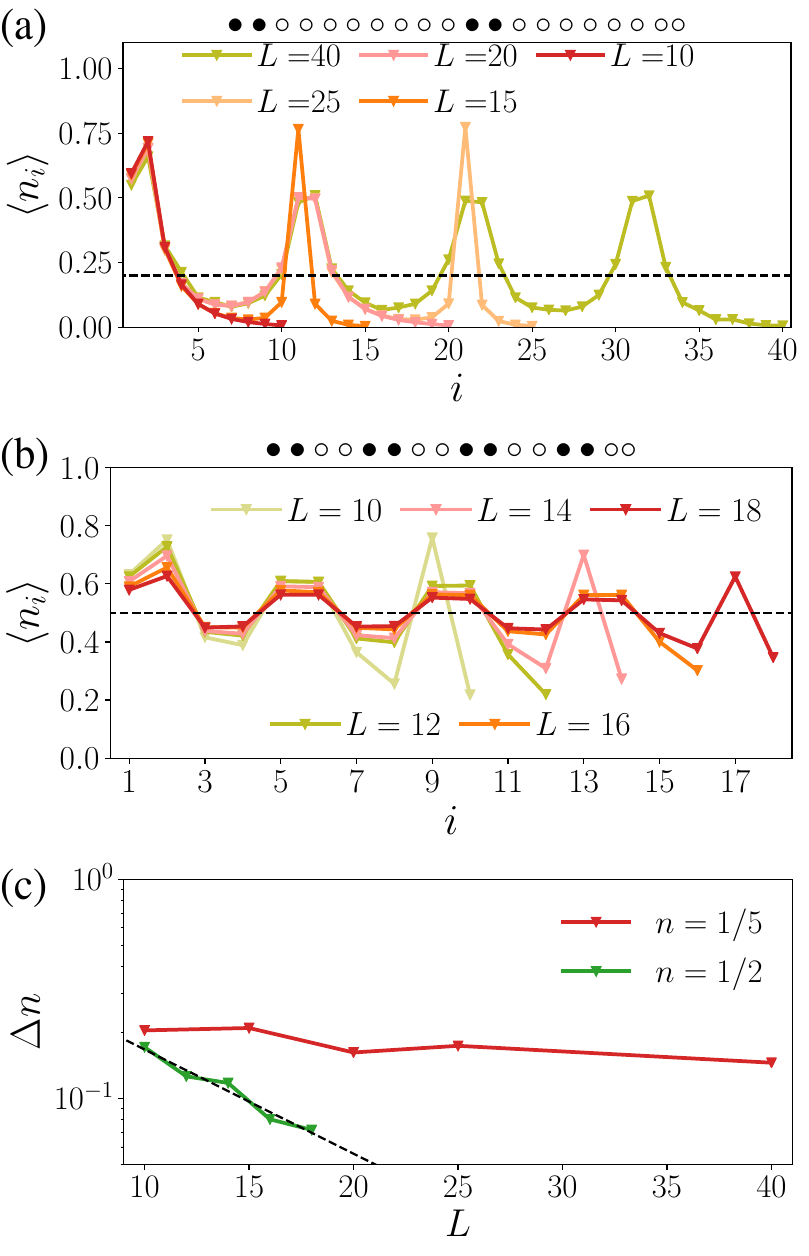}\\
\caption{The late time density profiles of the pair density waves at dense, $\nu=1/2$, and dilute, $\nu=1/5$, fillings show very different behavior. (a) Dilute configurations are essentially frozen, and do not approach the thermal density represented by the black dashed line.  (b) In contrast, at $\nu=1/2$ relaxation is enhanced at larger $L$.  (c) The deviation of late time density from the thermal value decay exponentially with system size as $e^{-L/\xi^\text{pair}_T}$ with $\xi^\text{pair}_T\approx 8.1$. In contrast for $\nu=1/5$, the residual density remains nearly constant with system size and na\"ive fit to the exponential gives an order of magnitude larger scale, $\xi^\text{dilute}_T\approx 84$.
Data at $\nu=1/5$ is obtained via ED ($L=10$, $L=15$, and $L=20$ with $5\times10^4$, $10^4$, and $2\times 10^3$ disorder realizations), Krylov time evolution ($L=25$, $T_\text{max}=10^3$ and $10^3$ disorder realizations) and TEBD ($L=40$, $T_\text{max}=300$ and $100$ disorder realizations). For $\nu=1/2$ we used ED ($L=10$, $12$, $14$, and $16$ with $3\times10^4$, $10^4$, $5\times10^3$, and $10^3$ disorder realizations) and Krylov time evolution ($L=18$, $T_\text{max}=10^3$, and $10^3$ disorder realizations). 
	\label{Fig:nPairs}}
\end{center}
\end{figure}

\begin{figure*}[t]
\begin{center}
\includegraphics[width=.45\columnwidth]{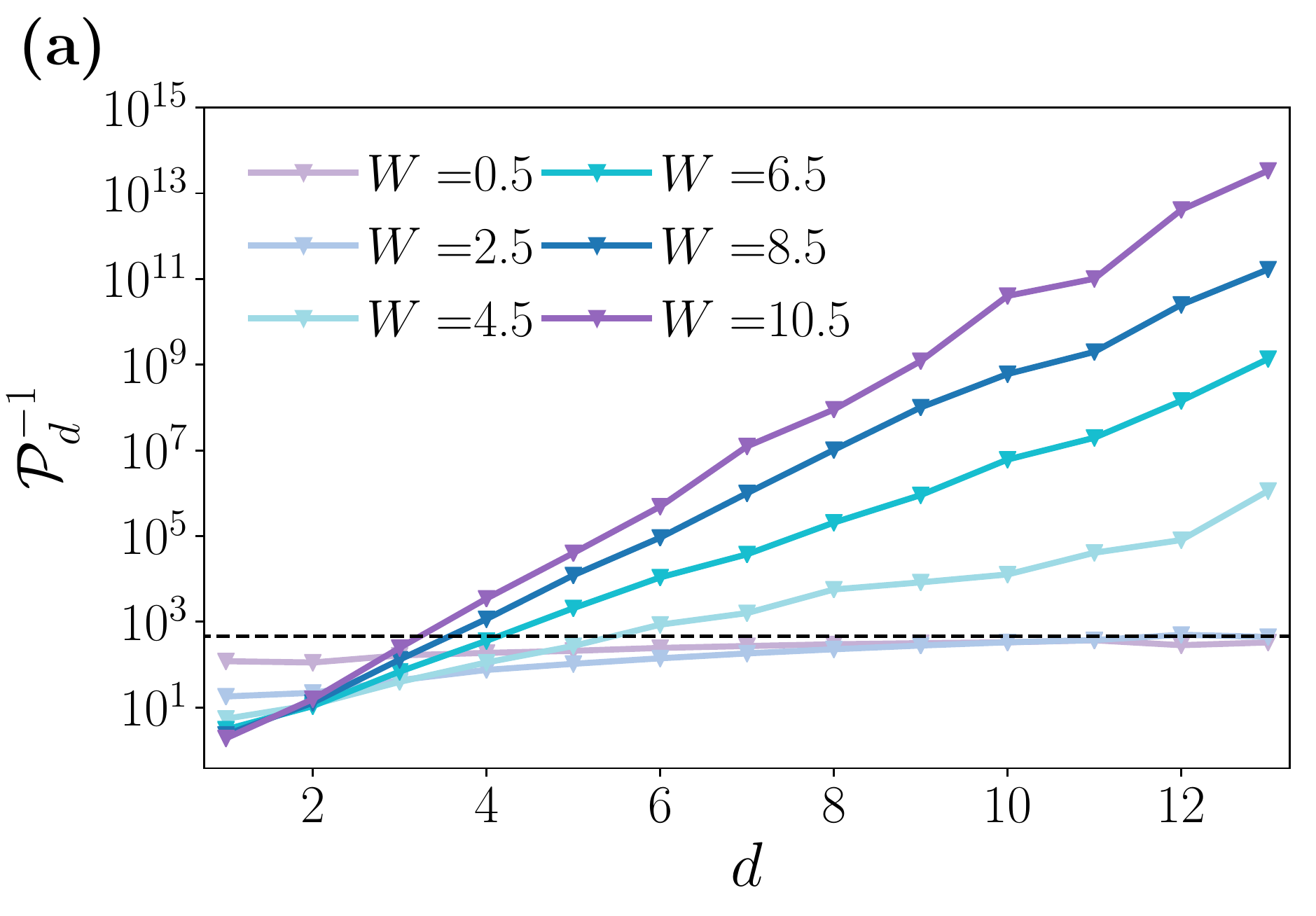}\quad \includegraphics[width=.45\columnwidth]{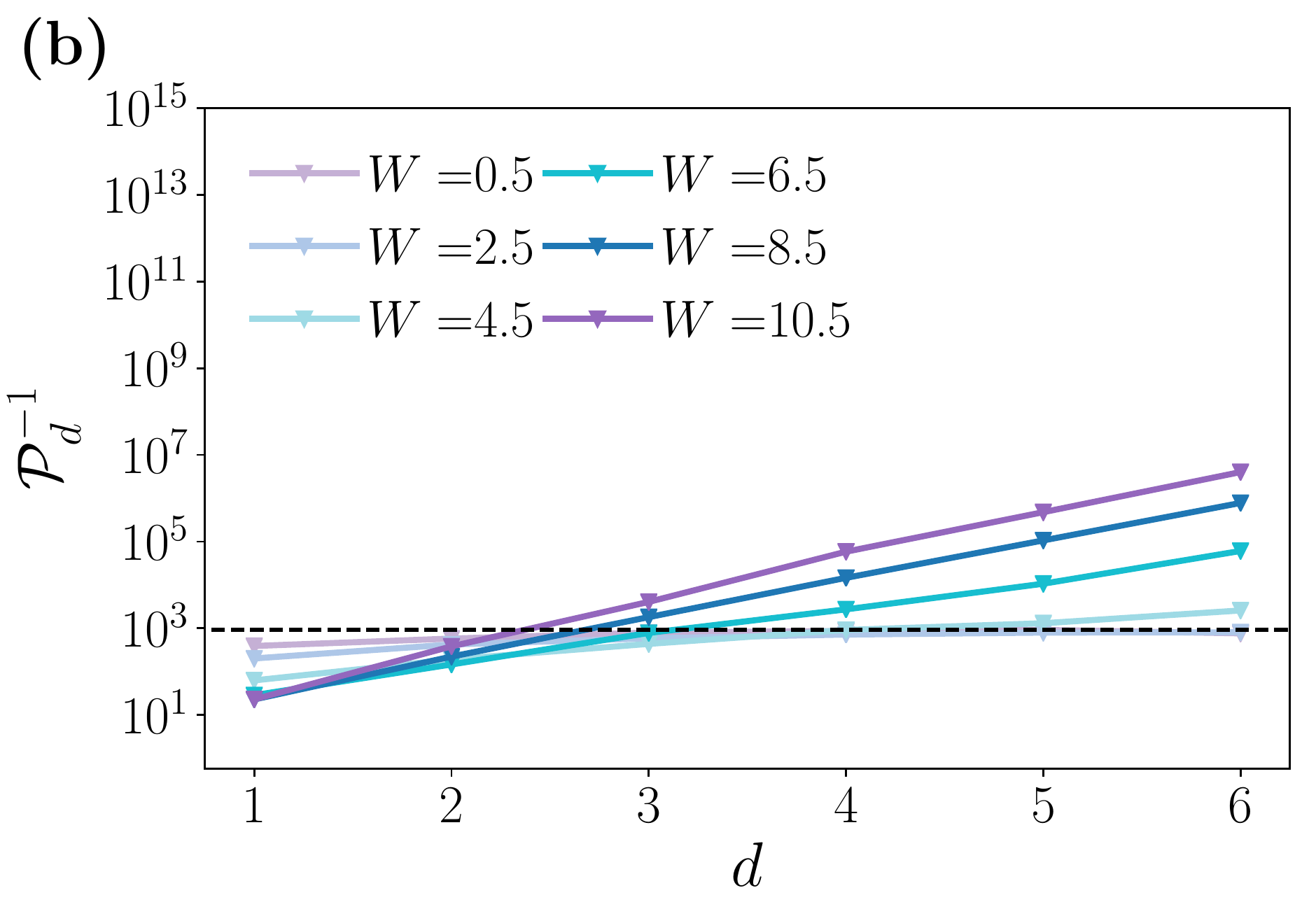}
\caption{The mutual IPR, ${\cal I}_d$, that quantifies the inverse probability of bubble tunneling $d$ sites, increases exponentially with $d$ at strong disorder. At weak disorder the mIPR approaches the Hilbert space dimension, $\mathcal{N}$, shown by a dashed line. In the dilute system in (b), $W=4.5$ marks the onset of the exponential growth, suggesting that the thermal bubble is frozen at its initial position. On the other hand, for $\nu=1/2$, in (a), the clear exponential behavior emerges only at larger disorder.  $\mathcal{I}_d$ was calculated for system sizes $L=15$ and $L=12$ in dilute and dense case respectively and averaged over $10^4$ disorder realizations. 
   \label{Fig:Mutual IPR-App} }
\end{center}
\end{figure*}

Below we consider quench from a pair-density wave of period $2/\nu$.  These configurations accommodate the maximal possible number of pairs in the uniform state. Figure~\ref{Fig:nPairs} confirms that such state is localized at $\nu=1/5$ and is relaxing in the dense case. Dense systems display strong dependence on the system size and increased tendency towards relaxation at larger system sizes, $L$. In contrast, at $\nu=1/5$ the late time density profile has almost no dependence on the size of the system. In particular, even at very large lengths the curves do not approach the average density represented by the dashed black line.

\section{$\mathcal{P}$ as a function of bubble distance}\label{Sec:MIPR}

In the text we introduced a quantity $\mathcal{P}(\ket{\psi_1},\ket{\psi_2})$ that measures how similar the expansion of $\psi_{1,2}$ over eigenstates is. We birefly mentioned that the case where $\ket{\psi_1}=\ket{\psi_2}$ corresponds to the usual participation ratio, hence we refer to the inverse of $\mathcal{P}$ as mutual inverse participation ratio (mIPR). The mutual IPR assumes very different values depending on the nature of the two states: values of mIPR  $O(\mathcal{N})$ correspond to two vectors that have similar expansion over eigenstates, while very large values of mIPR imply that the expansion is very different. In the main text we analyzed the mIPR between two product states where the bubble is located at the left and right end of the system respectively, see Eqs~(\ref{Eq:psi-bubble})-(\ref{Eq:psi-inverted}). Such pair of states corresponds to the maximum possible displacement of the bubble in the chain. Below we illustrate the behavior of mIPR between pair of states which correspond to a smaller bubble displacement. 

In our analysis we measure the mIPR, $\mathcal{P}^{-1}_d = \mathcal{P}^{-1}(\psi_L,\psi_d)$,  between the following states in the dense limit (half-filling, $L=12$),
\begin{align}
\label{Eq:psi d}
\ket{\psi_L} &= \boxed{\bullet\bullet\bullet\bullet\bullet\,\bullet}\circ\circ\circ\circ\circ\circ, \\
\ket{\psi_d} &= \underbrace{\circ\circ\circ}_d\boxed{\bullet\bullet\bullet\bullet\bullet\,\bullet}\circ\circ\circ.
\end{align}
Here we use the bubble that contains all particles to maximize the range of achievable displacements. For the dilute case, $L=15$, we use similar pair of states with bubble containing 3 particles ($\nu=1/5$).

In the thermal phase, eigenstates are approximately given by random vectors in the Hilbert space and their average overlap with other normalized vectors approaches the value predicted by random matrix theory, irrespective of the state or the eigenstate. In the weak disorder limit, we then expect $\mathcal{P}^{-1}_d$ to be independent on the distance between the two bubbles and to have the same behavior as the conventional IPR: $\mathcal{P}^{-1}_d\sim \mathcal{N}$. This expectation is confirmed by the results presented in fig.~\ref{Fig:Mutual IPR-App}(a) and (b) for $W=0.5$.

On the other hand, in the MBL phase eigenstates are not similar to random vectors, but instead are characterized by a set of local integrals of motion that have a finite overlap with the local particle density. Thus, two product states with globally different arrangement of particles are expected to have drastically different expansion over eigenstates. Therefore, we expect $\mathcal{P}^{-1}_d\propto\exp\left[d/\xi\right]$. As presented in figure~\ref{Fig:Mutual IPR-App}(a) and (b), at strong disorder our results support this hypothesis for both dilute (a) and dense (b) states. 

At intermediate disorder strength, we observe a qualitative difference between dense and dilute cases. Dilute configurations, Fig.~\ref{Fig:Mutual IPR-App}(a), show exponential behavior already at $W=4.5$, whereas dense states in Fig.~\ref{Fig:Mutual IPR-App}(b) need much stronger disorder to clearly present the same trend. This result confirms the presence of mobility edge and is consistent with the observed absence of pair spreading reported in Figure~\ref{Fig:nP} and also with the finite size scaling of mIPRs shown in the main text, Fig.~\ref{Fig:mutual-iprs}.

\section{Dynamical probe of the absence of resonances}

The discussion on mutual IPR showed how tunneling processes are strongly suppressed in the dilute case of our model. In addition to eigenstates analysis, we also studied long time dynamics of states with a thermal bubble. In this way, it was possible to verify whether a bubble initialized at a certain position can dynamically give rise to a dense region somewhere else in the chain. In order to study this process we defined a projector onto the subset of Hilbert space that has large density in a certain region. More specifically, we define
\begin{equation}
    \label{Eq:Pnc}
    \hat{P}_{\nu_c}(L_0,i)=\sum_{\ket{\phi_\alpha} \in \mathcal{C}}\ket{\phi_\alpha}\bra{\phi_\alpha},
\end{equation}
where states $\ket{\phi}$ are all possible product states that satisfy the condition $\nu\geq \nu_c$ in the region $[i,i+L_0]$. This projector selects all configurations where the system is locally above the mobility edge. We notice that $\hat{P}_{\nu_c}(L_0,i)$ takes into account all possible configurations, thus considering also the entropic factor. 

\begin{figure}[b]
    \begin{center}
    \includegraphics[width=.95\columnwidth]{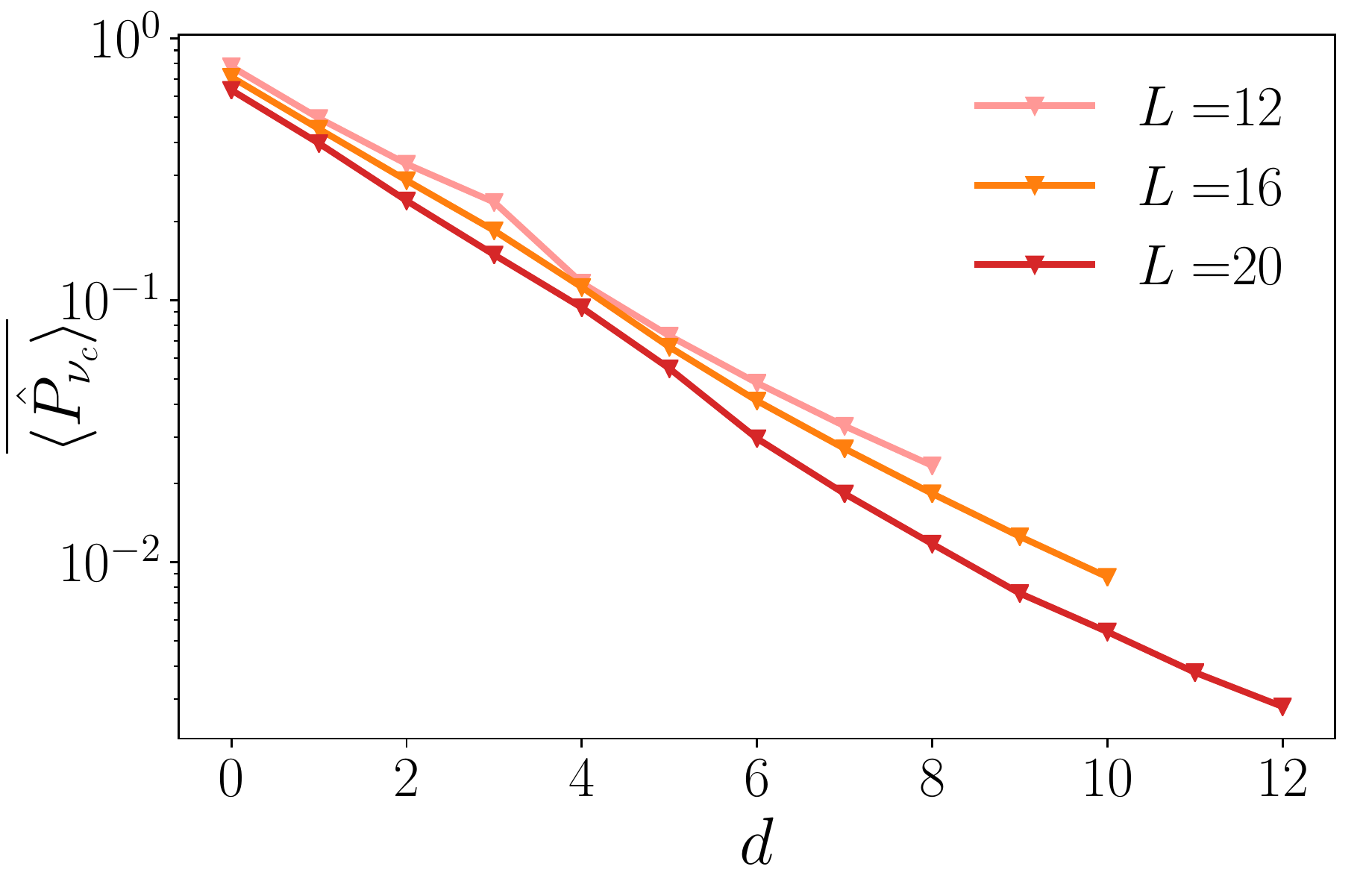}
    \caption{\label{Fig:Pnc} The late time evolution of $\overline{\langle{P}_{n_c}(L_0,d)\rangle}$ shows exponential decay for all the system sizes studied ($L=12,16,20$ at density $\nu=1/4$). Furthermore, we notice that increasing the system size the exponential vanishing becomes more severe, suggesting that in the thermodynamic limit there would be no motion of the bubble at all. These results were obtained using $10^4,5\times 10^3\text{ and } 10^3$ disorder realizations for the system sizes from smaller to larger. 
    }
    \end{center}
\end{figure}

 In order to understand what is the minimal required size of the region $L_0$, we use the lengthscale extracted from the decay of $\Delta n$. Fit in Fig.~\ref{Fig:quench}(c) in the main text yields $L_0\simeq 6\divisionsymbol 7$, while fit in Fig.~\ref{Fig:nPairs}(c) gives a somewhat larger scale. We define an initial state $\ket{\psi_0}$ that has an entangled dense region of size approximately $L_0$ (described by a linear superposition of product states $\ket{\phi_i}$) followed by a product state:
\begin{equation}
    \label{Eq:psi0 Pnc}
    \ket{\psi_0}=\frac{1}{\sqrt{N_\mathcal{C}}}\sum_{i=1}^{N_\mathcal{C}} \ket{\phi_i}\otimes \ket{\circ\circ\circ\bullet\circ\circ}.
\end{equation}
Below, we fix the overall density to $\nu=1/4$ and $W=6.5$, which still corresponds to a localized system. The dense region is obtained as a superposition of different configurations with $N-1$ particles in $L_0=2(N-1)$ sites. The remaining particle is initialized in the middle of the last segment of the chain. For instance, for $L=16$ this results into following initial state:
\begin{equation}
\label{Eq:psi Pnc}
\begin{split}
\ket{\psi_0}=\frac{1}{\sqrt{N_\mathcal{C}}}\Bigr[&\boxed{\bullet\bullet\circ\bullet\circ\circ}\circ\circ\circ\circ\bullet\circ\circ\circ\circ\circ \;+ \\ &\boxed{\bullet\circ\bullet\circ\bullet\circ}\circ\circ\circ\circ\bullet\circ\circ\circ\circ\circ \;+ \\ &\boxed{\bullet\bullet\bullet\circ\circ\circ}\circ\circ\circ\circ\bullet\circ\circ\circ\circ\circ \; +\dots \Bigr],
\end{split}
\end{equation}
where the boxed area contains a dense entangled bubble and the remainder is in the dilute state.

The initial state $\ket{\psi_0}$ is then evolved through the Hamiltonian~(\ref{Eq:H}) in a quench protocol. After time evolution up to a maximum time $T_\text{max}=1000$, we measure $\langle{P}_{\nu_c}(L_0,d)\rangle= \bra{\psi(t)}\hat{P}_{\nu_c}(L_0,d)\ket{\psi(t)}$, which quantifies the probability of encountering a bubble shifted by $d$ sites from the initial position of the bubble. 

Finally, averaging over all different product states in the dilute part of the chain and over disorder we obtain the data in Fig.~\ref{Fig:Pnc}. This plot reveals that the probability of having a dense ($\nu>\nu_c$) region decays exponentially with the distance $d$ from its initial location. This is in agreement with our long-time TEBD dynamics, Fig.~\ref{Fig:nP}, that reveals localization of individual pairs. Thus, we conclude that bubble does not spread resonantly but rather tunnels throughout the system. Moreover, the finite size scaling analysis shows that increasing the system size the decay of $\overline{\langle P_{\nu_c}(L_0,d)\rangle}$ with distance $d$ is enhanced. Therefore in the dilute regime of our model the bubble remains localized around its initial position.

\end{document}